\pdfoutput=1
\documentclass[12pt,a4paper]{article}

\usepackage{mathtools}
\usepackage{epsfig}
\usepackage{multirow}
\usepackage{cite}
\usepackage{mciteplus}

\usepackage{ifthen} 
\newboolean{pdflatex}
\setboolean{pdflatex}{true} 

\newboolean{articletitles}
\setboolean{articletitles}{true} 

\newboolean{uprightparticles}
\setboolean{uprightparticles}{false} 

\newcommand{\ups}{\ensuremath{\varUpsilon}}

\newcommand{\ones}{\ensuremath{\varUpsilon(1S)}}
\newcommand{\twos}{\ensuremath{\varUpsilon(2S)}}
\newcommand{\threes}{\ensuremath{\varUpsilon(3S)}}

\newcommand{\onesmm}{\ensuremath{\varUpsilon(1S)\rightarrow\mu^+\mu^-}}
\newcommand{\twosmm}{\ensuremath{\varUpsilon(2S)\rightarrow\mu^+\mu^-}}
\newcommand{\threesmm}{\ensuremath{\varUpsilon(3S)\rightarrow\mu^+\mu^-}}

\newcommand{\jpsi}{\ensuremath{$J/\psi$}}
\newcommand{\bit}{\begin{itemize}}
\newcommand{\bce}{\begin{center}}
\newcommand{\eit}{\end{itemize}}
\newcommand{\ece}{\end{center}}

\usepackage{booktabs} 
\usepackage{microtype}
\usepackage{lineno}  
\usepackage{xspace} 

\usepackage{graphicx}  
\usepackage{color}
\usepackage{colortbl}
\graphicspath{{./figs/}} 

\usepackage{amsmath} 
\usepackage{amssymb}
\usepackage{amsfonts}
\usepackage{upgreek} 

\newcommand*\patchAmsMathEnvironmentForLineno[1]{%
\expandafter\let\csname old#1\expandafter\endcsname\csname #1\endcsname
\expandafter\let\csname oldend#1\expandafter\endcsname\csname
end#1\endcsname
 \renewenvironment{#1}%
   {\linenomath\csname old#1\endcsname}%
   {\csname oldend#1\endcsname\endlinenomath}%
}
\newcommand*\patchBothAmsMathEnvironmentsForLineno[1]{%
  \patchAmsMathEnvironmentForLineno{#1}%
  \patchAmsMathEnvironmentForLineno{#1*}%
}
\AtBeginDocument{%
\patchBothAmsMathEnvironmentsForLineno{equation}%
\patchBothAmsMathEnvironmentsForLineno{align}%
\patchBothAmsMathEnvironmentsForLineno{flalign}%
\patchBothAmsMathEnvironmentsForLineno{alignat}%
\patchBothAmsMathEnvironmentsForLineno{gather}%
\patchBothAmsMathEnvironmentsForLineno{multline}%
}

\usepackage{hyperref}    
\usepackage[all]{hypcap} 

\def\sPlot{\mbox{\em sPlot}}
\def\lhcb {\mbox{LHCb}\xspace}
\def\ux85 {\mbox{UX85}\xspace}

\def\tevatron {Tevatron\xspace}

\ifthenelse{\boolean{uprightparticles}}%
{

 \def\Pmu         {\ensuremath{\upmu}\xspace}

 \def\Pchi        {\ensuremath{\upchi}\xspace}                 
 \def\Ppsi        {\ensuremath{\uppsi}\xspace}

 \def\PDelta      {\ensuremath{\Delta}\xspace}                 
 \def\PXi      {\ensuremath{\Xi}\xspace}                 
 \def\PLambda      {\ensuremath{\Lambda}\xspace}                 
 \def\PSigma      {\ensuremath{\Sigma}\xspace}                 
 \def\POmega      {\ensuremath{\Omega}\xspace}                 
 \def\PUpsilon      {\ensuremath{\Upsilon}\xspace}                 
 
 \def\PB      {\ensuremath{\mathrm{B}}\xspace}                 
                  
 \def\PD      {\ensuremath{\mathrm{D}}\xspace}

 \def\PJ      {\ensuremath{\mathrm{J}}\xspace}                 
 \def\PK      {\ensuremath{\mathrm{K}}\xspace}

 \def\Pb      {\ensuremath{\mathrm{b}}\xspace}                 
 \def\Pc      {\ensuremath{\mathrm{c}}\xspace}

 \def\Pi      {\ensuremath{\mathrm{i}}\xspace}

}
{

 \def\Pmu         {\ensuremath{\mu}\xspace}

 \def\Pchi        {\ensuremath{\chi}\xspace}                 
 \def\Ppsi        {\ensuremath{\psi}\xspace}                 
                  
 \mathchardef\PDelta="7101
 \mathchardef\PXi="7104
 \mathchardef\PLambda="7103
 \mathchardef\PSigma="7106
 \mathchardef\POmega="710A
 \mathchardef\PUpsilon="7107
                  
 \def\PB      {\ensuremath{B}\xspace}                 
                  
 \def\PD      {\ensuremath{D}\xspace}

 \def\PJ      {\ensuremath{J}\xspace}                 
 \def\PK      {\ensuremath{K}\xspace}

 \def\Pb      {\ensuremath{b}\xspace}                 
 \def\Pc      {\ensuremath{c}\xspace}

 \def\Pi      {\ensuremath{i}\xspace}

}

\def\mumu       {\ensuremath{\Pmu^+\Pmu^-}\xspace}

\def\cquark    {\ensuremath{\Pc}\xspace}

\def\bquark    {\ensuremath{\Pb}\xspace}
\def\bquarkbar {\ensuremath{\overline \bquark}\xspace}

\def\kaon  {\ensuremath{\PK}\xspace}
  \def\Kbar  {\kern 0.2em\overline{\kern -0.2em \PK}{}\xspace}

\def\Kz    {\ensuremath{\kaon^0}\xspace}
\def\Kzb   {\ensuremath{\Kbar^0}\xspace}
\def\KzKzb {\ensuremath{\Kz \kern -0.16em \Kzb}\xspace}
\def\Kp    {\ensuremath{\kaon^+}\xspace}
\def\Km    {\ensuremath{\kaon^-}\xspace}

\def\KpKm  {\ensuremath{\Kp \kern -0.16em \Km}\xspace}

  \def\Dbar    {\kern 0.2em\overline{\kern -0.2em \PD}{}\xspace}
\def\D       {\ensuremath{\PD}\xspace}

\def\Dz      {\ensuremath{\D^0}\xspace}
\def\Dzb     {\ensuremath{\Dbar^0}\xspace}
\def\DzDzb   {\ensuremath{\Dz {\kern -0.16em \Dzb}}\xspace}
\def\Dp      {\ensuremath{\D^+}\xspace}
\def\Dm      {\ensuremath{\D^-}\xspace}

\def\DpDm    {\ensuremath{\Dp {\kern -0.16em \Dm}}\xspace}

  \def\Bbar    {\kern 0.18em\overline{\kern -0.18em \PB}{}\xspace}

\def\jpsi     {\ensuremath{{\PJ\mskip -3mu/\mskip -2mu\Ppsi\mskip 2mu}}\xspace}

  \def\Y#1S{\ensuremath{\PUpsilon{(#1S)}}\xspace}

\def\chic  {\ensuremath{\Pchi_{c}}\xspace}

\def\Lbar {\ensuremath{\kern 0.1em\overline{\kern -0.1em\PLambda}}\xspace}

\def\to                 {\ensuremath{\rightarrow}\xspace}

\def\AT#1     {\ensuremath{A_{\mathrm{T}}^{#1}}\xspace}           

\def\C#1      {\ensuremath{\mathcal{C}_{#1}}\xspace}                       
\def\Cp#1     {\ensuremath{\mathcal{C}_{#1}^{'}}\xspace}                    
\def\Ceff#1   {\ensuremath{\mathcal{C}_{#1}^{\mathrm{(eff)}}}\xspace}        
\def\Cpeff#1  {\ensuremath{\mathcal{C}_{#1}^{'\mathrm{(eff)}}}\xspace}       
\def\Ope#1    {\ensuremath{\mathcal{O}_{#1}}\xspace}                       
\def\Opep#1   {\ensuremath{\mathcal{O}_{#1}^{'}}\xspace}                    

\newcommand{\tev}{\ensuremath{\mathrm{\,Te\kern -0.1em V}}\xspace}
\newcommand{\gev}{\ensuremath{\mathrm{\,Ge\kern -0.1em V}}\xspace}
\newcommand{\mev}{\ensuremath{\mathrm{\,Me\kern -0.1em V}}\xspace}
\newcommand{\kev}{\ensuremath{\mathrm{\,ke\kern -0.1em V}}\xspace}
\newcommand{\ev}{\ensuremath{\mathrm{\,e\kern -0.1em V}}\xspace}
\newcommand{\gevc}{\ensuremath{{\mathrm{\,Ge\kern -0.1em V\!/}c}}\xspace}
\newcommand{\mevc}{\ensuremath{{\mathrm{\,Me\kern -0.1em V\!/}c}}\xspace}
\newcommand{\gevcc}{\ensuremath{{\mathrm{\,Ge\kern -0.1em V\!/}c^2}}\xspace}
\newcommand{\gevgevcccc}{\ensuremath{{\mathrm{\,Ge\kern -0.1em V^2\!/}c^4}}\xspace}
\newcommand{\mevcc}{\ensuremath{{\mathrm{\,Me\kern -0.1em V\!/}c^2}}\xspace}

\def\mm   {\ensuremath{\rm \,mm}\xspace}

\def\mum  {\ensuremath{\,\upmu\rm m}\xspace}

\def\invpb {\ensuremath{\mbox{\,pb}^{-1}}\xspace}

\def\ps   {\ensuremath{{\rm \,ps}}\xspace}

\newcommand{\chisq}{\ensuremath{\chi^2}\xspace}

\def\gsim{{~\raise.15em\hbox{$>$}\kern-.85em
          \lower.35em\hbox{$\sim$}~}\xspace}
\def\lsim{{~\raise.15em\hbox{$<$}\kern-.85em
          \lower.35em\hbox{$\sim$}~}\xspace}

\def\sPlot{\mbox{\em sPlot}}

\def\pt         {\mbox{$p_{\rm T}$}\xspace}

\newcommand{\lum} {\ensuremath{\mathcal{L}}\xspace}

\def\evtgen     {\mbox{\textsc{EvtGen}}\xspace}
\def\pythia     {\mbox{\textsc{Pythia}}\xspace}

\def\geant      {\mbox{\textsc{Geant4}}\xspace}

\def\photos     {\mbox{\textsc{Photos}}\xspace}

\def\tell1  {TELL1\xspace}
\def\ukl1   {UKL1\xspace}

\usepackage{cite} 
\usepackage{mciteplus}

\begin{document}

\renewcommand{\thefootnote}{\fnsymbol{footnote}}
\setcounter{footnote}{1}


\begin{titlepage}
\pagenumbering{roman}

\vspace*{-1.5cm}
\centerline{\large EUROPEAN ORGANIZATION FOR NUCLEAR RESEARCH (CERN)}
\vspace*{1.5cm}
\hspace*{-0.5cm}
\begin{tabular*}{\linewidth}{lc@{\extracolsep{\fill}}r}
\ifthenelse{\boolean{pdflatex}}
{\vspace*{-2.7cm}\mbox{\!\!\!\includegraphics[width=.14\textwidth]{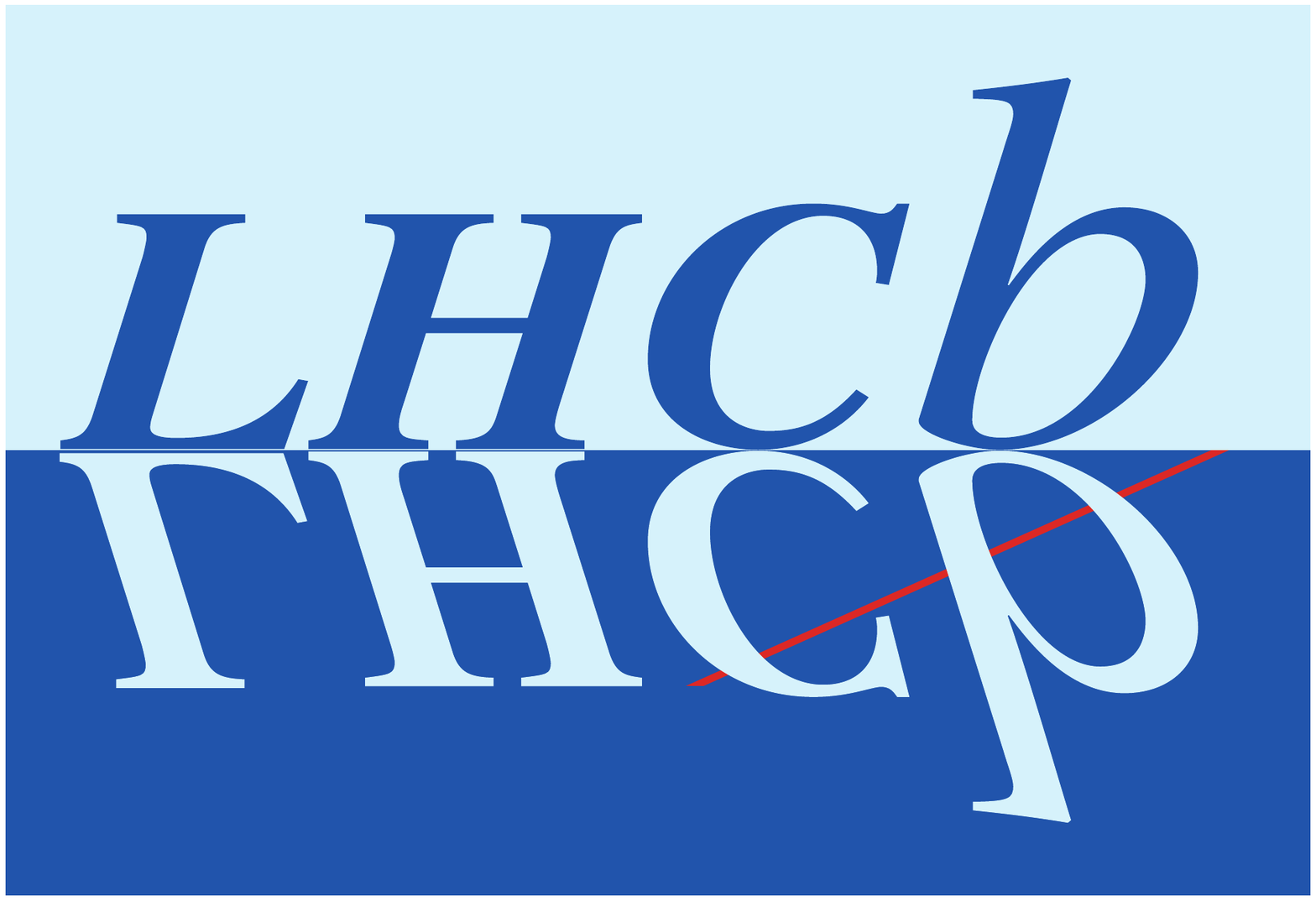}} & &}%
{\vspace*{-1.2cm}\mbox{\!\!\!\includegraphics[width=.12\textwidth]{lhcb-logo.eps}} & &}%
\\
 & & CERN-PH-EP-2013-071 \\  
 & & LHCb-PAPER-2013-016 \\  
 & & 22$^{nd}$ June 2013 \\ 
\end{tabular*}

\vspace*{3.0cm}

{\bf\boldmath\huge
\begin{center}
 Production of \jpsi~and $\ups$ mesons in $pp$ collisions at $\sqrt{s}=8~\tev$
\end{center}
}

\vspace*{1.5cm}

\begin{center}
The \lhcb collaboration\footnote{Authors are listed on the following pages.}
\end{center}

\vspace{\fill}

\begin{abstract}
  \noindent
The production of \jpsi\ and $\ups$~mesons in $pp$ collisions at $\sqrt{s}=8~\tev$  is studied with the LHCb detector.
The \jpsi\ and $\ups$~mesons are reconstructed in the $\mu^+\mu^-$ 
decay mode and the signal yields are determined with a fit to the  $\mu^+\mu^-$ invariant mass distributions. 
The analysis is performed in the rapidity range $2.0<y<4.5$ and transverse momentum range $0<\pt<14\,(15)~\gevc$ 
of the $\jpsi(\ups$) mesons. The $\jpsi$ and $\ups$ production cross-sections and 
the fraction of $\jpsi$ mesons from $b$-hadron decays are measured as a function 
of the meson \pt and $y$.
\end{abstract}
\vspace*{2.0cm}

\begin{center}
  Submitted to JHEP
\end{center}

{\footnotesize {\footnotesize
\centerline{\copyright~CERN on behalf of the \lhcb collaboration,
license \href{http://creativecommons.org/licenses/by/3.0/}{CC-BY-3.0}.}}}
\vspace*{2mm}

\vspace{\fill}

\end{titlepage}


\newpage
\setcounter{page}{2}
\mbox{~}
\newpage

\centerline{\large\bf LHCb collaboration}
\begin{flushleft}
\small
R.~Aaij$^{40}$, 
C.~Abellan~Beteta$^{35,n}$, 
B.~Adeva$^{36}$, 
M.~Adinolfi$^{45}$, 
C.~Adrover$^{6}$, 
A.~Affolder$^{51}$, 
Z.~Ajaltouni$^{5}$, 
J.~Albrecht$^{9}$, 
F.~Alessio$^{37}$, 
M.~Alexander$^{50}$
% ...

{\footnotesize \it
% ...
$ ^{5}$Clermont Universit\'{e}, Universit\'{e} Blaise Pascal, CNRS/IN2P3, LPC, Clermont-Ferrand, France\\
$ ^{6}$CPPM, Aix-Marseille Universit\'{e}, CNRS/IN2P3, Marseille, France\\
% ...
$ ^{9}$Fakult\"{a}t Physik, Technische Universit\"{a}t Dortmund, Dortmund, Germany\\
% ...
$ ^{35}$Universitat de Barcelona, Barcelona, Spain\\
$ ^{36}$Universidad de Santiago de Compostela, Santiago de Compostela, Spain\\
$ ^{37}$European Organization for Nuclear Research (CERN), Geneva, Switzerland\\
% ...
$ ^{40}$Nikhef National Institute for Subatomic Physics, Amsterdam, The Netherlands\\
% ...
$ ^{45}$H.H. Wills Physics Laboratory, University of Bristol, Bristol, United Kingdom\\
% ...
$ ^{50}$School of Physics and Astronomy, University of Glasgow, Glasgow, United Kingdom\\
$ ^{51}$Oliver Lodge Laboratory, University of Liverpool, Liverpool, United Kingdom\\
% ...
$ ^{n}$LIFAELS, La Salle, Universitat Ramon Llull, Barcelona, Spain\\
% ...
$ ^{s}$Scuola Normale Superiore, Pisa, Italy\\
}
\end{flushleft}

\cleardoublepage

\renewcommand{\thefootnote}{\arabic{footnote}}
\setcounter{footnote}{0}


\pagestyle{plain} 
\setcounter{page}{1}
\pagenumbering{arabic}

\newcommand{\prompt}{\ensuremath{\mathrm{prompt}~\jpsi}}
\newcommand{\fromb}{\ensuremath{\jpsi~\mathrm{from}~\bquark}}
\newcommand{\tpm}{$ & $\,\pm\,$ & $}
 
\section{Introduction} 
\label{sec:Introduction}
Successfully describing heavy quarkonium production 
is a long-standing problem in QCD.  An~effective field theory, non-relativistic QCD (NRQCD)~\cite{CaswellLepage1986PL,PhysRevD.51.1125},
provides the foundation for much of the current theoretical work. 
According to NRQCD, the production of heavy quarkonium 
factorises into two steps:  a heavy quark-antiquark pair is first created 
at short distances and subsequently evolves
non-perturbatively into quarkonium at long distances. 
The NRQCD calculations depend on the colour-singlet (CS) and colour-octet (CO) matrix elements, 
which account for the probability of a heavy quark-antiquark pair 
in a particular colour state to evolve into a heavy quarkonium state. 
The CS model (CSM)~\cite{Kartvelishvili:1978id,Baier:1981uk}, which provides
a leading-order description of quarkonium production, was initially used 
to describe experimental data. However, it underestimates the~observed
cross-section for single \jpsi~production at high transverse momentum (\pt) at the 
\tevatron~\cite{Abe:1992ww}. To~resolve this
discrepancy, the CO mechanism was introduced~\cite{Braaten:1994vv}. 
The corresponding matrix elements were determined from the
high-\pt data, as the CO cross-section decreases  more slowly with \pt
than that predicted by CS. More recent higher-order
calculations~\cite{Campbell:2007ws,Gong:2008sn,artoisenet:2008,lansberg:2009}
close the gap between the CS predictions and the experimental data~\cite{Brambilla:2010cs},
reducing the need for large CO contributions.

Studies of the production of the $\jpsi$ and $\ones$, $\twos$ and $\threes$ 
mesons (indicated generically as $\ups$ in the following)   have been performed
using $pp$ collision data taken at 
$\sqrt{s} = 7~\tev$ and at $\sqrt{s} = 2.76~\tev$
by the LHCb~\cite{LHCb-PAPER-2011-003,LHCB-PAPER-2011-036,LHCb-PAPER-2012-039},
 ALICE~\cite{AliceJpsi,AliceJpsi2, AliceJpsi3}, ATLAS~\cite{AtlasJpsi,AtlasUpsilon} 
and CMS~\cite{CmsJpsi,CmsUpsilon,CMSUpsilon2} experiments in 
different kinematic regions.
As well as providing direct tests of the underlying
production mechanism, these studies are crucial to estimate the
contribution of double parton scattering to multiple quarkonium
production~\cite{LHCb-PAPER-2011-013,LHCB-PAPER-2012-003}. 

In this paper first measurements of quarkonium production at $\sqrt{s}=8~\tev$ 
are reported under the assumption of  zero polarisation, an assumption that 
is discussed in the paper.
The differential production cross-sections of prompt \jpsi\ and \ups\ mesons, 
produced at the $pp$ collision point either directly or via feed-down from higher mass
charmonium or bottomonium states, are presented in the range of 
rapidity $2.0<y<4.5$ and $\pt<14\,\gevc$ ($\jpsi$) or $\pt<15\,\gevc$ ($\ups$). 
The fraction of \jpsi\  mesons from $b$-hadron decays, abbreviated as ``\fromb'' in the following, is also 
measured in the same fiducial region.

\section{The LHCb detector and data set}
The \lhcb detector~\cite{Alves:2008zz} is a single-arm forward
spectrometer covering the \mbox{pseudorapidity} range $2<\eta <5$,
designed for the study of particles containing \bquark or \cquark
quarks. The detector includes a high precision tracking system
consisting of a silicon-strip vertex detector surrounding the $pp$
interaction region, a large-area silicon-strip detector located
upstream of a dipole magnet with a bending power of about
$4{\rm\,Tm}$, and three stations of silicon-strip detectors and straw
drift tubes placed downstream. The combined tracking system has a
momentum resolution $\Delta p/p$ that varies from 0.4\% at 5\gevc to
0.6\% at 100\gevc, and an impact parameter resolution of 20\mum for
tracks with high \pt. Charged hadrons are identified
using two ring-imaging Cherenkov detectors~\cite{LHCb-DP-2012-003}. Photon, electron and
hadron candidates are identified by a calorimeter system consisting of
scintillating-pad and preshower detectors, an electromagnetic
calorimeter and a hadronic calorimeter. Muons are identified by a
system composed of alternating layers of iron and multiwire
proportional chambers, with the exception of the centre of the first station, which uses triple-GEM detectors.

The data sample used in this analysis was collected during the first part of the data taking period at $\sqrt{s}=8~\tev$ 
in April 2012.   During this period the average number of interactions per crossing varied.  The $\ups$ meson analysis 
is based on a data sample, corresponding to an integrated luminosity of about $51~\invpb$ of $pp$ interactions, collected 
with an average of 1.3  visible interactions per crossing. The analysis for the more abundant  \jpsi\ mesons is based on data, corresponding to an 
 integrated luminosity of about $18~\invpb$, collected with an average of 1.0 visible interactions per crossing. %
The trigger~\cite{LHCb-DP-2012-004} consists of a hardware stage, based on information from the calorimeter and muon
systems, followed by a software stage, which applies a full event reconstruction.
At the hardware stage, events are selected requiring dimuon candidates  with a product of their \pt
 larger than $1.68\,(\gevc)^2$. In the subsequent software trigger, two well reconstructed tracks are required to have hits in the muon 
system,  a \pt higher than $500\,\mevc$, $p$ higher than $6\,\gevc$ and to form a common vertex.  Only events with a dimuon candidate 
with an invariant mass $m_{\mu\mu}$ within $120\,\mevcc$ of the known $\jpsi$ meson
mass~\cite{PDG2012} or  larger than  $4.7\,\gevcc$ are retained for further analysis.  

\section{Selection and cross-section determination} 
\label{sec:cross}
The selection is based on the criteria described in Refs.~\cite{LHCb-PAPER-2011-003,LHCB-PAPER-2011-036} and is summarised in Table~\ref{selection}. 
It starts by combining oppositely-charged particles, identified as muons, with a track \pt larger than 700\,(1000)$\,\mevcc$ 
for the $\jpsi\,\!(\ups)$\,meson.  Good track quality is ensured by requiring  a $\chi^2$ per degree of freedom, $\chi^2/{\rm ndf}$, 
less than 4 for the track fit.
Duplicate particles created by the reconstruction are suppressed to the level of  $0.5 \times 10^{-3}$ using  the 
 Kullback-Leibler (KL) distance variable~\cite{kl,*kl2,*kl3}. To ensure good quality vertex reconstruction, the $\chi^2$  probability of the dimuon vertex is required to be larger than $0.5\,\%$. 
In addition, the primary vertex (PV) associated to the dimuon candidate is required to 
be within the luminous region, defined as  $|x_{\rm PV}| <1\,\mm$ , $|y_{\rm PV}| <1\,\mm$ and $|z_{\rm PV}| <150\,\mm$.

In the \jpsi\ analysis additional criteria are applied to the vertex quality. The  uncertainty on the pseudo decay time $t_z$, defined in Eq.~\ref{eq:tz}, is required to be less than $0.3\,\ps$, as estimated by the propagation
 of the uncertainties given by the track reconstruction. 
\begin{table}[!tb]
\caption{\small Selection criteria for the \jpsi\ and $\ups$ meson analyses. Criteria common to both analyses are 
displayed between the two columns.}\label{tab::cuts}
\begin{center}
\begin{tabular}{lcc} 
\midrule
Variable & Value ($\jpsi$)  & Value ($\ups$)  \\ \midrule
Track $\pt$ (\mevc)   & $> 700$ &   $>1000$ \\
Track $ \chi^2/{\rm ndf}$ & \multicolumn{2}{c}{$<~4$} \\
KL distance  &  \multicolumn{2}{c}{$> 5000$} \\
Vertex $\chi^2$ probability & \multicolumn{2}{c}{ $>0.5 \%$} \\
$t_z$ uncertainty~({\rm ps}) & $< 0.3$ & --  \\  
Mass window $m_{\mu\mu}$\,(\mevcc)& $|m_{\mu\mu} - M(\jpsi)| < 120$ & $8500<m_{\mu\mu}<11500$\\
\bottomrule
\end{tabular}
\label{selection}
\end{center}
\end{table}

The simulation samples are based on the \pythia 6.4 generator~\cite{Sjostrand:2006za} configured with the parameters detailed in 
Ref.~\cite{LHCb-PROC-2010-056}. 
The \evtgen package~\cite{Lange:2001uf} is used to generate hadron decays.
The interaction of the generated particles with the detector and its
response are implemented using the \geant toolkit~\cite{Allison:2006ve, *Agostinelli:2002hh} as described in
Ref.~\cite{LHCb-PROC-2011-006}.
Radiative corrections to the decay of the vector meson to dimuons are generated with
the \photos package~\cite{Golonka:2005pn}. 

The differential cross-section  for the production of a vector meson $V$ in a  bin of ($\pt,y$), where $V$ stands 
for a \jpsi\ or $\ups$ meson, decaying into a muon pair, is 
\begin{equation}
\frac{{\rm d}^2\sigma}{{\rm d}y{\rm d}\pt} (pp\to VX) = \frac{N\left(V\to\mu^+\mu^-\right)}{\lum \times \epsilon_{\rm tot} \times {\cal B}\left(V\to\mu^+\mu^-\right)\times \Delta y \times \Delta \pt}\label{eq:sigma}\,,
\end{equation}
where $N\left(V \to \mu^+ \mu^-\right)$ is the number of observed $V \to \mu^+ \mu^-$ candidates, $\epsilon_{\rm tot}$
the total detection efficiency 
in the given bin, \lum is the integrated luminosity, ${\cal B}\left(V\to\mu^+\mu^-\right)$  is the branching fraction of the $V
\to \mu^+\mu^-$ decay and $\Delta y=0.5$ and $\Delta \pt=1\gevc$
are the rapidity and \pt\ bin sizes, respectively. In the case of
the $J/\psi \to \mu^+ \mu^- $ decay the branching fraction is well known, ${\cal B}(\jpsi\ \to\mumu)  = ( 5.94 \pm 0.06 )\times 10^{-2}$~\cite{PDG2012}, and therefore it is chosen to quote an
absolute cross-section. On the other hand,  the dimuon branching fractions of the $\ups$ mesons are known less precisely\cite{PDG2012}, and
therefore, as in Ref.~\cite{LHCB-PAPER-2011-036}, the product of the cross-section times the dimuon branching fraction is given.

The total efficiency $\epsilon_{\rm tot}$ 
is the product of the
geometric acceptance, the reconstruction and selection efficiency and the trigger efficiency.
All  efficiency terms are evaluated using simulated samples and validated with data-driven techniques in each  ($\pt,y$) bin.

The procedure to measure the integrated luminosity is described in Ref.~\cite{LHCb-PAPER-2011-015}. For this analysis a van der Meer scan~\cite{VanDerMeer} was performed in April 2012, resulting in a measurement of the integrated luminosity of $18.4\pm0.9\,\invpb$ for the \jpsi\ and  $50.6\pm 2.5\,\invpb$ for the $\ups$ samples.

\section{{\boldmath \jpsi} meson signal}
\label{sec:jpsisig}

As in the previous studies, prompt \jpsi\ mesons are distinguished from \fromb\ by means of the pseudo decay time variable
 defined as
\begin{equation}\label{eq:tz}
t_z = \frac{(z_{\jpsi}-z_{\rm PV}) \times M_{\jpsi}}{p_z}\,,
\end{equation}
where $z_{\jpsi}$ and $z_{\rm PV}$ are the positions along the beam axis $z$ of the \jpsi decay vertex and of the primary vertex
refitted after removing the decay muons of the \jpsi candidate; $p_z$ is the measured \jpsi momentum in the beam direction and $M_{\jpsi}$ is the known \jpsi mass~\cite{PDG2012}. 

The yields of both prompt \jpsi\ mesons and \fromb\ are determined from a two-dimensional  fit in each ($\pt,y$) bin to the distributions of invariant mass and pseudo decay time  of the signal candidates, following the approach described in Ref.~\cite{LHCb-PAPER-2011-003}. The mass distribution is modelled with a Crystal Ball function~\cite{Skwarnicki:1986xj} for the signal and an exponential function for the combinatorial background. 

\begin{figure}[t!]
\begin{center}
\begin{tabular}{cc}
\raisebox{0.0\height}{\includegraphics[width=0.5\textwidth]{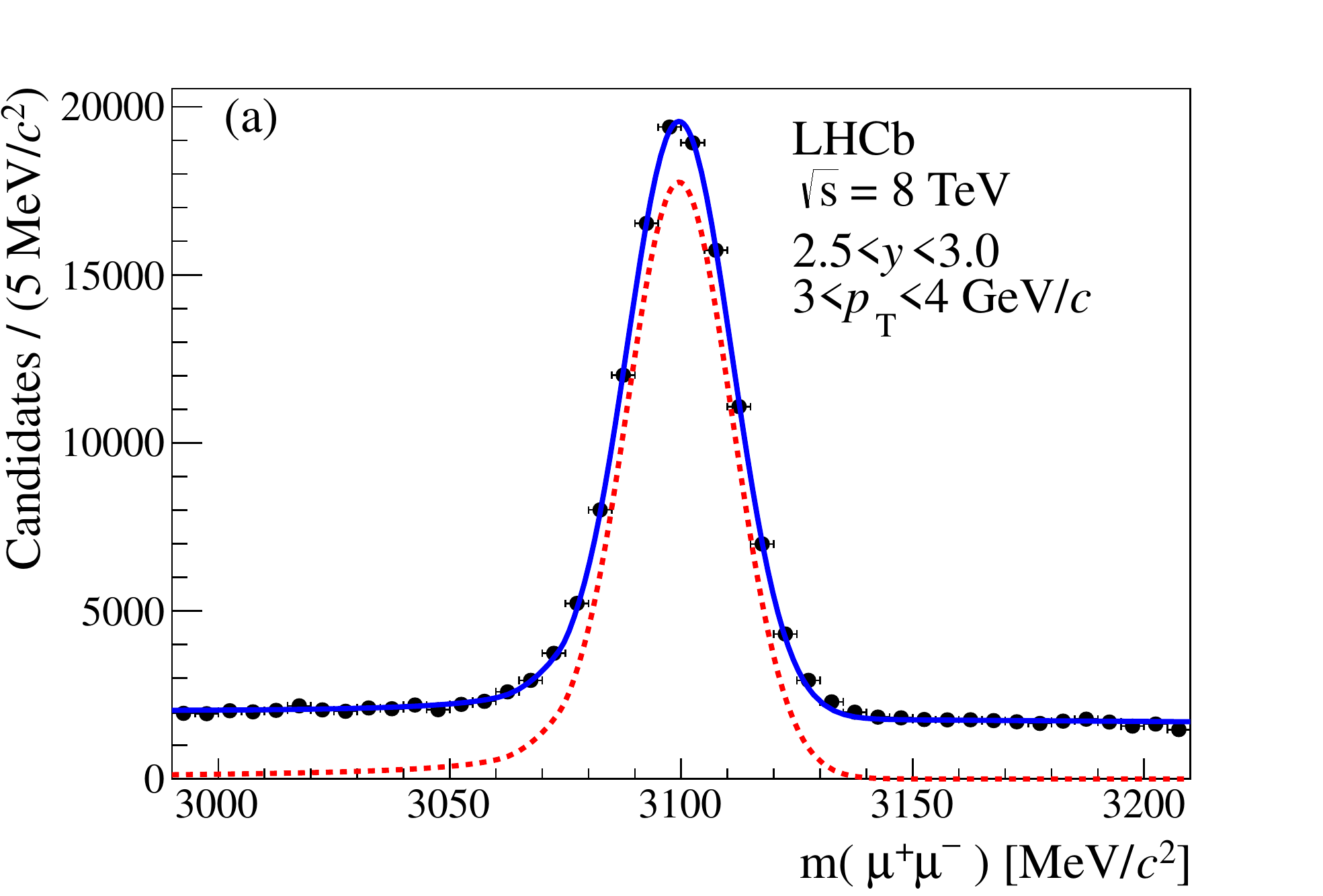}}& 
\includegraphics[width=0.5\textwidth]{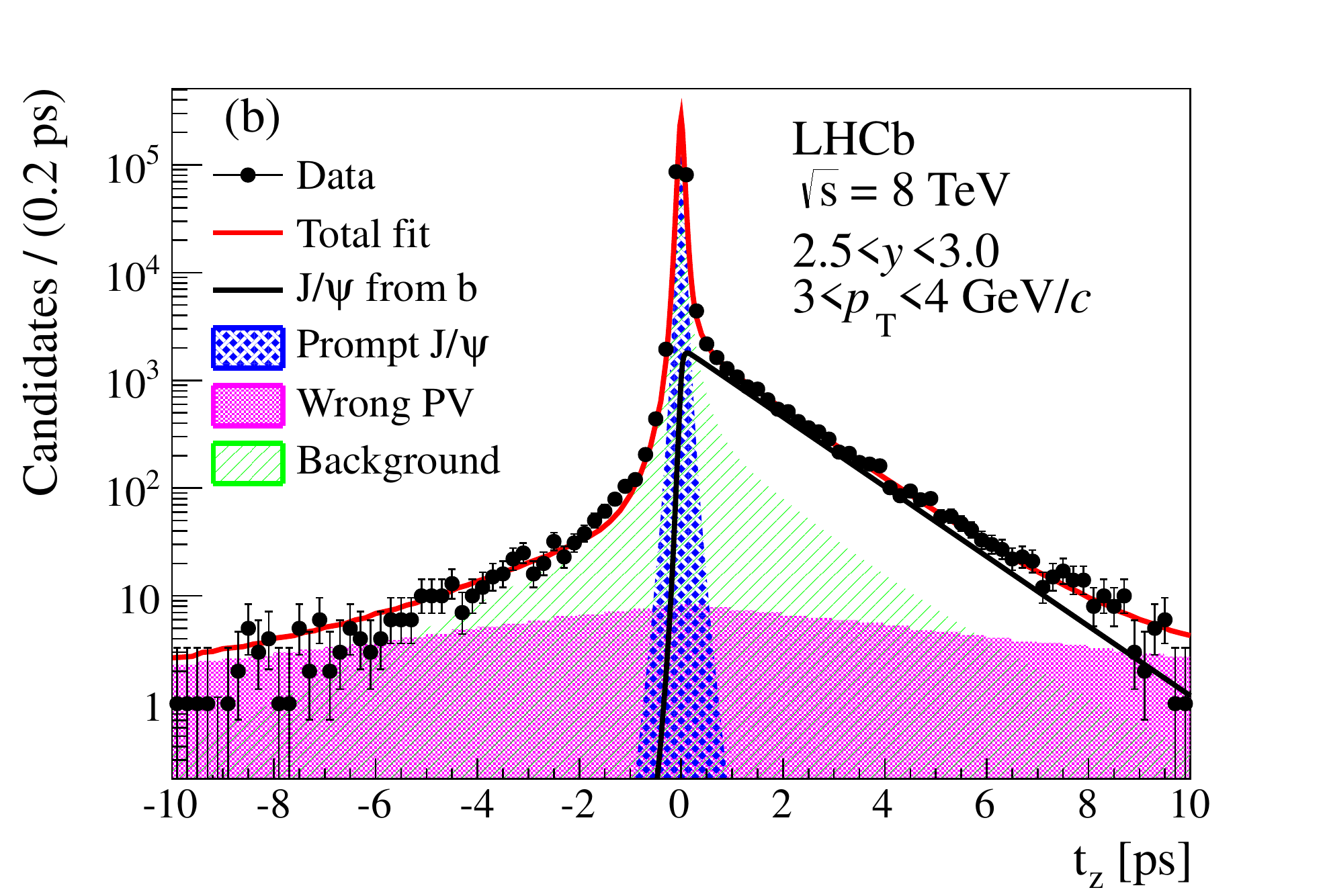} \\
\end{tabular}
\caption{\small Projections of the fit result for a selected bin in \pt and $y$
  for (a) the $\jpsi$ invariant dimuon mass and (b) $t_z$. For the
  former, the total fitted function is shown (blue solid line) together
  with the signal  distribution (red dotted line). 
  In the $t_z$ projection the total fitted function is shown together with the \fromb\ component, the prompt signal, the background and  the tail component due to the association of a \jpsi\ candidate with a wrong PV. }
\label{exampletime}
\end{center}
\end{figure}

The signal pseudo decay time distribution is described by a $\delta$-function at $t_z=0$ for the prompt \jpsi\ component together with an exponential decay function for the \fromb\ component. 
The shape of the tail arising from the association  of a \jpsi\ meson candidate with a wrong primary vertex is derived from the data by combining a \jpsi\ meson from a given event with the primary vertex of the following event in the sample. The prompt component of the signal function and that from $b$ hadron decays  are convolved with  a resolution function modelled by the  sum of two Gaussian functions. The background distribution is parameterised with an empirical function based on the shape of the $t_z$ distribution observed in the \jpsi\ mass sidebands. It is 
built as the sum of a 
$\delta$-function and five exponential components, three for positive $t_z$ and two for negative $t_z$, 
the negative and positive 
exponential functions with the largest lifetime having their lifetimes $\tau_{\rm L}$ fixed to the same value.  
This function is convolved with  a resolution function modelled by the  sum of two Gaussian functions. 
All parameters of the background component are determined independently in each ($\pt,y$) bin from the distribution of the pseudo decay time and are fixed in the final fit. The total fit function is the sum of the products of the mass and $t_z$ fit functions for the signal and background. Figure~\ref{exampletime} shows the fit projections in  mass and  $t_z$  for one specific bin ($3 < \pt <4\,\gevc$, \,$2.5<y<3.0$) with the fit result superimposed.  Summing over all bins, a total signal yield of 2.6 million $\jpsi$ events is obtained.

\section{ {\boldmath\ups} meson signal}

The $\ups$ meson signal yields are determined from a fit to the reconstructed dimuon invariant mass of the selected candidates 
with $8.5<m_{\mu\mu}<11.5~\gevcc$. The distribution is described by the sum of three Crystal Ball functions, one for each of 
 the $\ones$, $\twos$ and $\threes$ signals, and an exponential function for the combinatorial background. The 
parameters $\alpha$ and $n$ 
of the Crystal Ball function describing the radiative tail are fixed to the values of $\alpha=2$ and $n=1$ based on 
simulation studies. The width of the Crystal Ball 
function describing the $\ones$ meson is allowed to vary, while the widths of the $\twos$ and $\threes$  mesons are constrained to 
the value of the width of the 
$\ones$ signal, scaled by the ratio of the masses of the $\twos$ and $\threes$ to the $\ones$ meson. The peak values of  
the $\ones$, $\twos$ and $\threes$ mass distributions are allowed to vary in the fit and are consistent with the known 
values~\cite{PDG2012}.
\begin{figure}[t!]
\bce
 \includegraphics[width=0.5\textwidth]{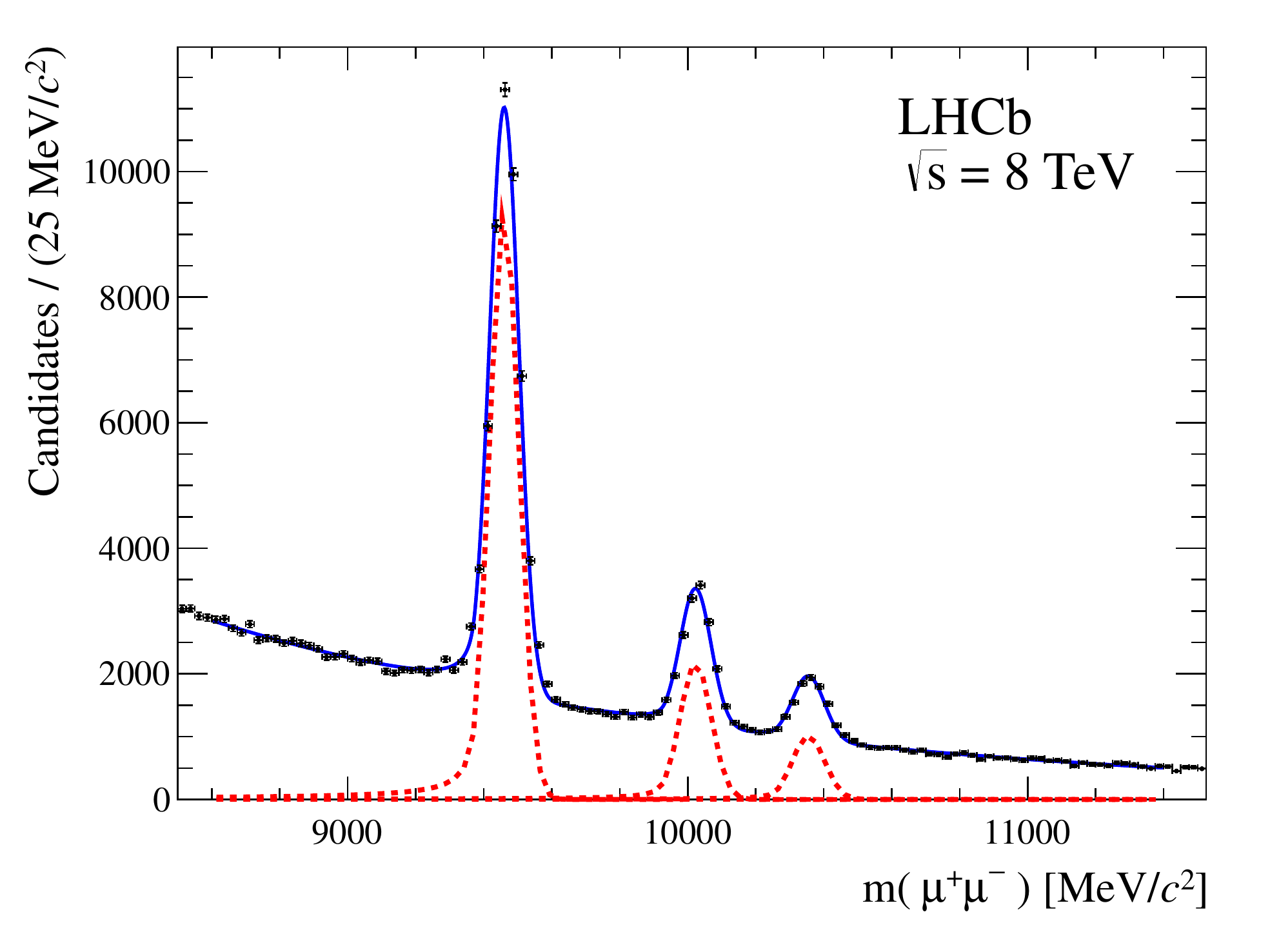}
 \label{fig2}
 \caption{\small Invariant mass distribution of the selected $\ups\to\mu^+\mu^-$ candidates in the range $\pt<15\,\gevc$ and
$2.0<y<4.5$. The three peaks correspond to the $\ones$, $\twos$ and $\threes$ meson signals (from left to right). 
The superimposed curve and the signal yields (dotted) are the result of the fit described in the text. }
\ece
\end{figure}
Figure~\ref{fig2} shows the results of the fit performed over the entire range in $\pt$ and $y$.
The obtained signal yields are $43\,785\pm254$, $10\,976\pm155$ and $5325\pm122$ for the 
$\ones$, $\twos$ and $\threes$ mesons, respectively,  with a mass resolution of the $\ones$ resonance of $43\mevcc$.
The fit is repeated independently for each of the bins in $\pt$ and $y$. When fitting the individual bins, the masses are fixed  to the values obtained when fitting the full range, while the mass resolution for the $\ones$ candidates is parameterised with a linear function of $\pt$ and $y$ that reproduces the behaviour observed in data.
Bins with too few entries are excluded from the analysis.

\section{Systematic uncertainties}\label{sec:systematics}

Previous studies~\cite{LHCb-PAPER-2011-003,LHCB-PAPER-2011-036} have shown that the total efficiency depends on the initial 
polarisation state of the vector meson. 
The $\jpsi$ polarisation has been measured at $\sqrt{s}=7~\tev$ 
by the LHCb~\cite{LHCb-PAPER-2013-008} and ALICE~\cite{AliceJpsiPola} collaborations, 
in a kinematic range similar to that used in this analysis, and the 
$\ups$ polarisation has been measured by CMS~\cite{CMSUpsilonPola} 
at large $\pt$ and central rapidity.
They were both found to be small.
Therefore, in this paper results are quoted under the assumption of zero 
polarisation and no corresponding systematic uncertainty is assigned
on the cross-section for this effect.
All other systematic uncertainties are summarised in Table~\ref{syst}.
\renewcommand{\arraystretch}{1.2}
\begin{table}[!tb]
\tabcolsep 4mm
\begin{center}
\caption{\small \label{syst}Relative systematic uncertainties (in \%) on the $\jpsi$ and $\ups$ 
cross-section results and on the fraction
of \fromb.}
\begin{tabular}{@{}llll@{}}
\midrule
\multicolumn{3}{l}{\it Correlated between bins} \\ \midrule
& Mass fits                  & \multicolumn{2}{c}{$\phantom{1}0.7\ {\rm to}\ 2.2$} \\
& Radiative tail            & \multicolumn{2}{c}{$\phantom{1}1.0$}\\
& Muon identification       & \multicolumn{2}{c}{$\phantom{1}1.3$}\\
& Tracking efficiency       & \multicolumn{2}{c}{$\phantom{1}0.9$}\\
& Vertexing                 & \multicolumn{2}{c}{$\phantom{1}1.0$}\\
& Trigger                   & \multicolumn{2}{c}{$\phantom{1}4.0$}\\
& Luminosity                & \multicolumn{2}{c}{$\phantom{1}5.0$}\\
& ${\cal B}(\jpsi\to\mumu)$ & \multicolumn{2}{c} {$\phantom{1}1.0$} \\
\midrule
\multicolumn{3}{l}{\it Uncorrelated between bins} \\ \midrule
 & Production model  & \multicolumn{2}{c} {1.0\ {\rm to}\ 6.0} \\

& $t_z$ fit, for \jpsi from $b$ & \multicolumn{2}{c} {1.0\ {\rm to}\ 12.0} \\
\bottomrule
\end{tabular}
\end{center}

\end{table}

Uncertainties related to the mass model describing the shape of the
dimuon mass distribution are estimated by fitting the invariant mass distributions for the $\jpsi$ and
$\ups$ mesons  with the sum of two Crystal Ball functions. 
The relative difference in the number of 
signal events (0.7--2.2\%) is taken as a systematic uncertainty. 
A fraction of events has a lower invariant mass because of the
energy lost through bremsstrahlung.
Based on simulation studies, about 4\% of the signal
events are estimated to be outside the analysis mass windows and are
not counted as signal. The fitted signal yields are corrected for this
effect and an uncertainty of 1\% is assigned to the cross-section
measurement based on a comparison between the radiative tail observed
in data and simulation.

The uncertainty due to the muon identification efficiency is measured on data using a tag-and-probe method. This 
method reconstructs \jpsi\ candidates in which one muon is identified by the muon system (``tag'') and the other 
(``probe'') is identified selecting a track depositing the energy of minimum-ionising particles in the calorimeters.

The ratio of the muon identification efficiency measured in data to that obtained in the simulation is convolved 
with the momentum distribution of muons from 
\jpsi and \ups\ mesons to obtain an efficiency correction. This is found to be $0.98\pm0.01$; the uncertainty on 
the correction factor is considered as a systematic uncertainty.

The uncertainty on the reconstruction efficiency of the muon tracks has also been estimated using a data-driven tag-and-probe approach 
based on partially reconstructed  \jpsi decays, and it was found to be 0.9\% per muon pair. 

Differences between data and simulation 
in the efficiency of the requirement on the vector meson vertex \chisq probability lead to a further uncertainty of 1\%. 

The trigger efficiency is determined using a data-driven method exploiting a sample of events that are still triggered when the signal candidate is removed \cite{LHCb-DP-2012-004}. The efficiency obtained with this method in each ($\pt,y$) bin is used to check the efficiencies measured in the simulation. 
The systematic uncertainty associated with the trigger efficiency is the difference between that measured in the data and in the simulation. As a cross-check, the trigger efficiency is also computed using  a data sample that has not been required to pass any physics trigger. The results obtained with the two methods are consistent.

The luminosity is determined with an uncertainty of 5\%, dominated by differences in the results obtained with a van der Meer scan~\cite{VanDerMeer} using the core and off-core parts of the beam. 

The dependence of the efficiency calculation on the production model used in the simulation is taken into account by varying the main parameters of the \pythia 6.4 generator
related to prompt vector meson production. 
These parameters define the minimum \pt cut-offs for regularising the cross-section. This effect is evaluated  in each ($\pt,y$) bin and found to be at most 6\%.

Uncertainties related to the $t_z$ fitting procedure for the \jpsi\ mesons
are included by changing the parameterisation used to describe the signal and  background. 
A second fitting method  based on the \sPlot\ technique~\cite{Pivk:2004ty} is used with the mass as the control variable  to unfold the background and to perform an unbinned likelihood fit to the pseudo decay time distribution. The two approaches give consistent results and their difference  is taken as an estimate of the systematic uncertainty. These uncertainties are evaluated  in each ($\pt,y$) bin and found to be a few percent. 

\section{Results on {\boldmath\jpsi} meson production}
\label{jpsiresults}
\begin{figure}[b!]
\begin{center}
\includegraphics[width=0.6\textwidth]{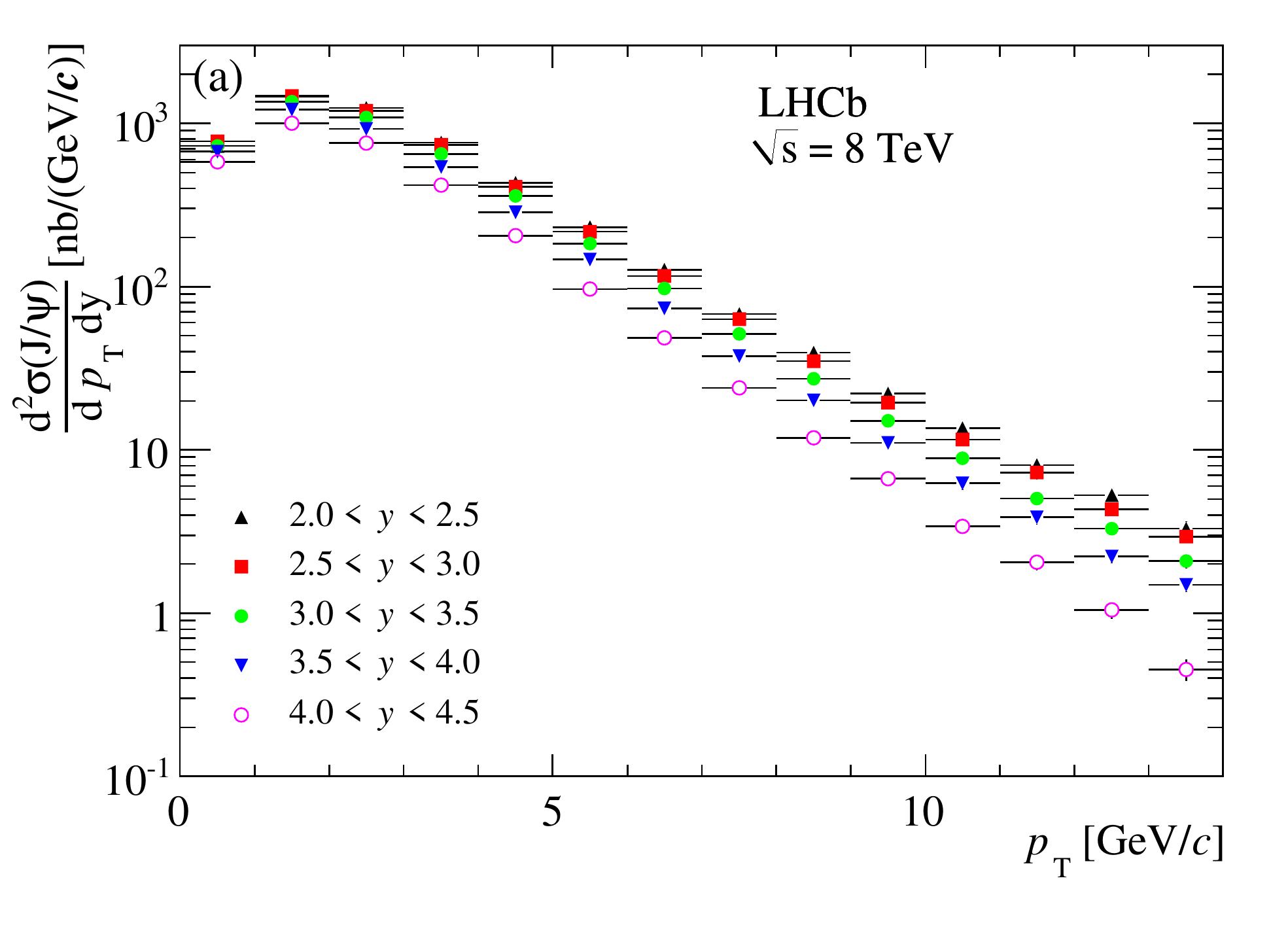}
\includegraphics[width=0.6\textwidth]{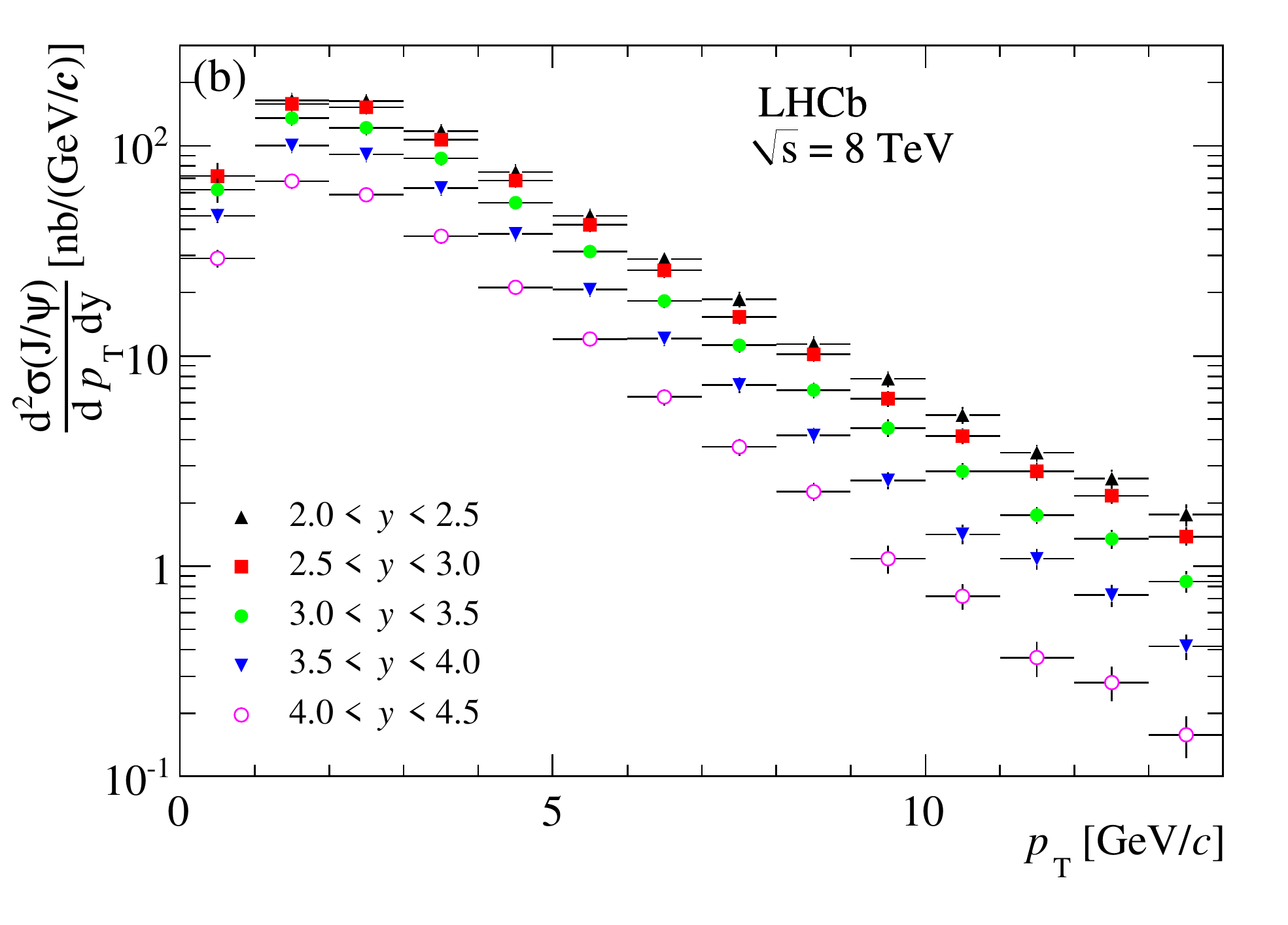}
\caption{\small Differential production cross-section for (a) prompt \jpsi\ mesons
   and (b) \fromb\  as a function of \pt in bins of $y$. It is
  assumed that prompt \jpsi\ mesons are produced unpolarised. The errors are the quadratic sums of the statistical and
systematic uncertainties.}
\label{fig:xsjpsiprompt}
\end{center}
\end{figure}

The measured  double-differential production cross-sections for  prompt \jpsi\ mesons, under the assumption of zero polarisation, and for \fromb\  are given in bins of \pt\ and $y$ in Tables~\ref{promptresult} and~\ref{bresult}, respectively, and are displayed in Fig.~\ref{fig:xsjpsiprompt}.  
The integrated cross-section for prompt \jpsi\ meson production in the
defined fiducial region, summing over all bins of the analysis, is
\begin{equation*}
\sigma\left({\rm prompt}\,\, \jpsi, \pt <14\,\gevc,\,2.0<y<4.5\right)  \, = \, 
10.94\pm 0.02\pm 0.79\,\upmu{\rm b},
\end{equation*}
where the first uncertainty is statistical and the second is systematic, computed taking correlations into account. The integrated cross-section for the production of  \fromb\ in the same fiducial region is
\begin{equation*}\label{sigma_from_b}
\sigma\left(\fromb,\, \pt <14\,\gevc,\,2.0<y<4.5\right)  \, =  \, 1.28 \pm 0.01\pm 0.11\,\upmu{\rm b}.
\end{equation*} 

The total $\bquark\bquarkbar$ production cross-section is computed as 
\begin{equation}
\sigma(pp \to b\overline{b} X) = 
\alpha_{4\pi} \,\frac{\sigma\left(\fromb, \,\pt <14\,\gevc,\,2.0<y<4.5\right)}{2 \, {\cal B}(b\to\jpsi X)},
\end{equation}
where the factor $\alpha_{4\pi}=5.4$ is an extrapolation factor of the cross-section from the measured
to the full kinematic region. This factor is obtained using the simulation as described in Sect.~\ref{sec:cross}.
The inclusive $b{\to}\jpsi X$ branching fraction is \mbox{${\cal B}(b\to\jpsi X)=(1.16\pm0.10)\%$}~\cite{PDG2012}. 
The resulting total $b\overline{b}$ cross-section is \mbox{$\sigma(pp \to b\overline{b} X) = 298\pm 2 \pm 36\, \upmu{\rm b}$}, where the first uncertainty is statistical and the second is systematic, which includes the uncertainty on ${\cal B}(b\to\jpsi X)$.
No systematic uncertainty has been included for the extrapolation factor $\alpha_{4\pi}$ estimated from the simulation. For comparison, the value of the
extrapolation factor given by NLO calculations is 5.1~\cite{Cacciari}.

\begin{figure}[!b]
\centering
\includegraphics[width=0.48\textwidth]{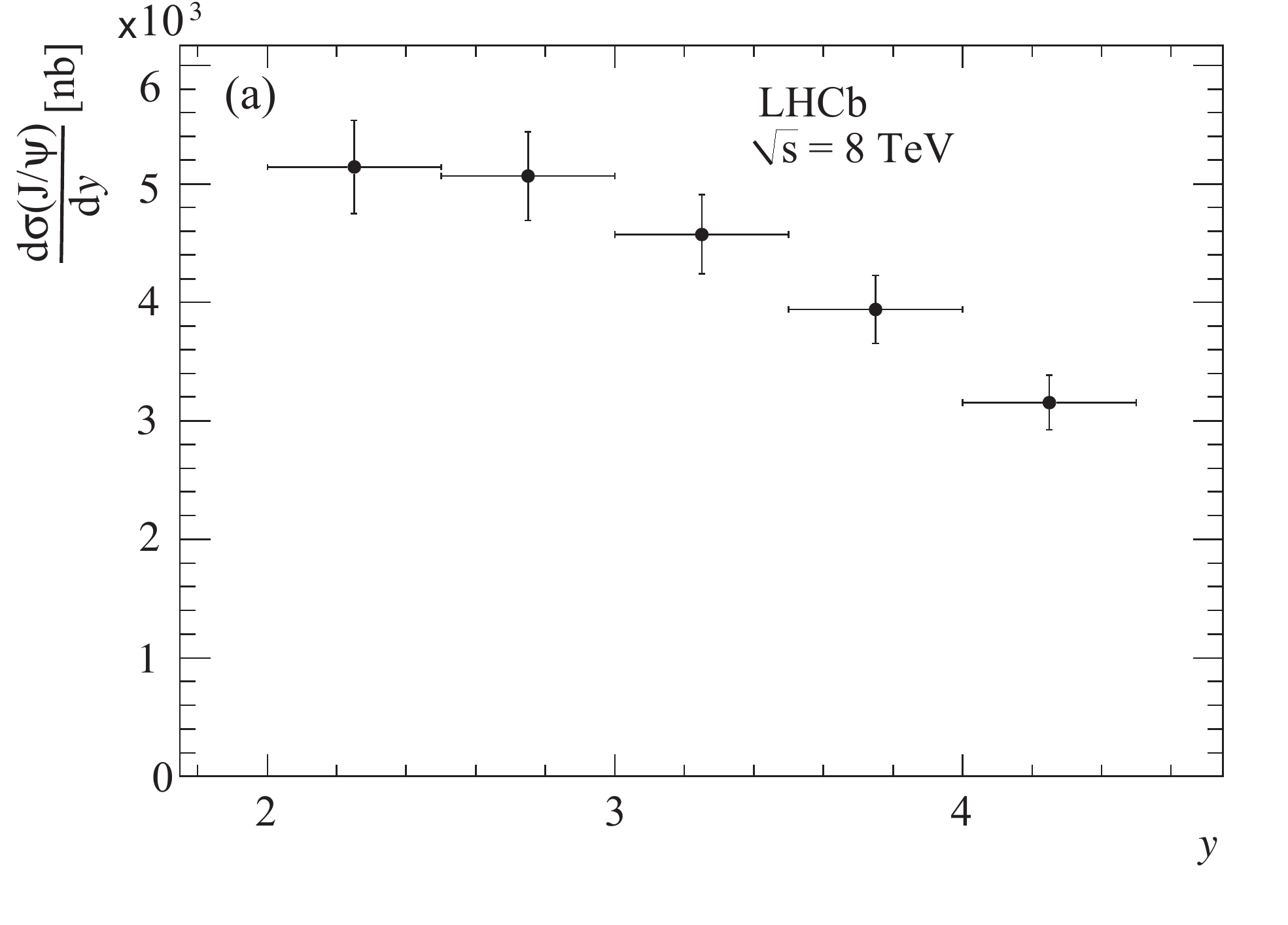}
\includegraphics[width=0.48\textwidth]{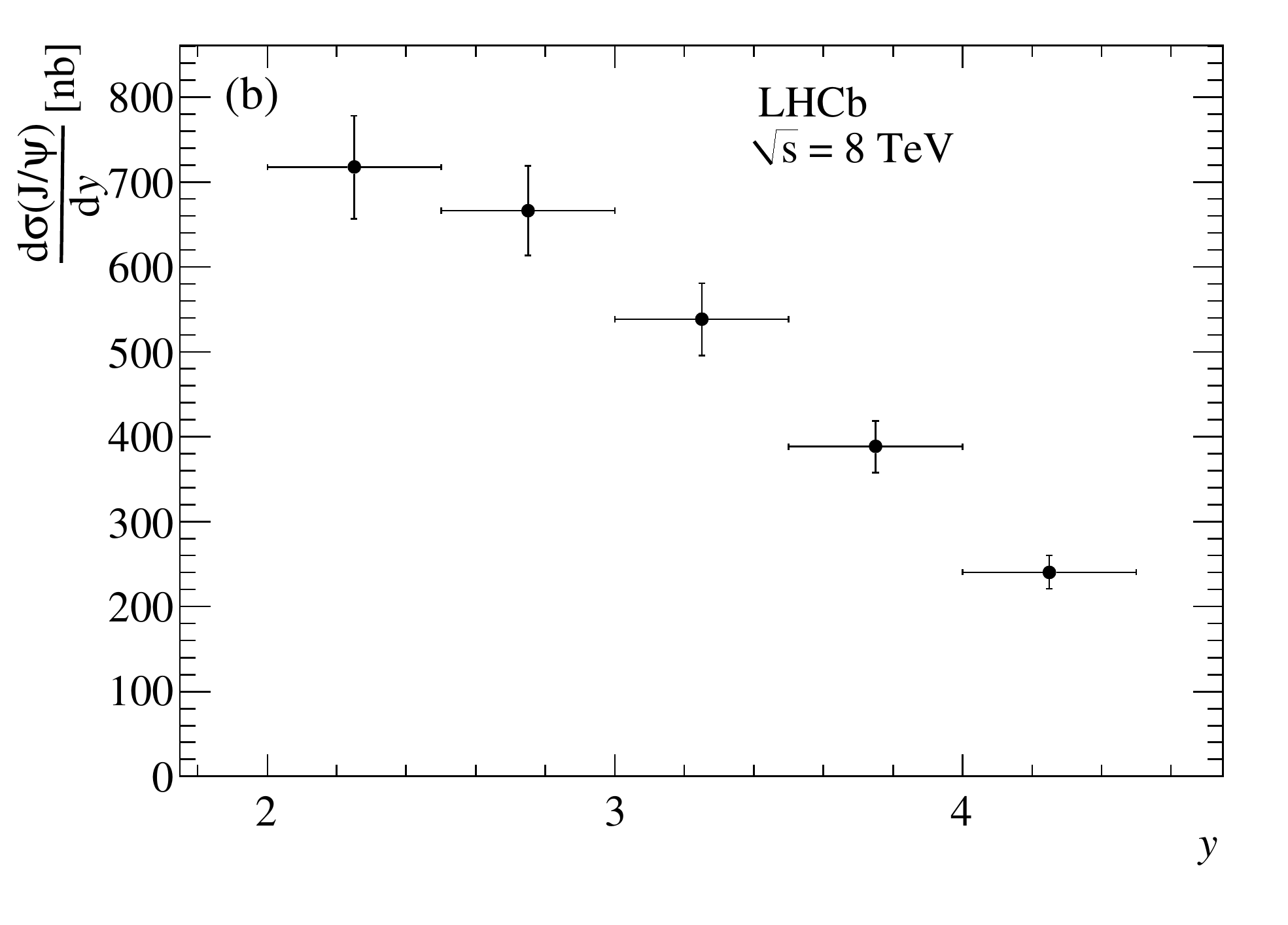}
\caption{\small Differential production cross-section as a function of $y$ integrated over \pt, for (a) unpolarised
\prompt\  mesons and (b) \fromb. The errors are the quadratic sums of the statistical and 
systematic uncertainties.} \label{fig:rapidity}
\end{figure}

Table~\ref{tab:rapidity}  and Fig.~\ref{fig:rapidity}  show the differential production cross-section ${\rm d}\sigma/{\rm d}y$ 
integrated over \pt,  for unpolarised \prompt\ mesons and \fromb. For both components, the cross-section decreases significantly between the central and forward regions of the acceptance.
\renewcommand{\arraystretch}{1.3}
\begin{table}[!t]
\caption{\small Differential production cross-section ${\rm d}\sigma/{\rm d}y$ in nb for \prompt\ mesons (assumed unpolarised) and for \fromb, integrated 
over \pt. The first uncertainty is statistical, the second (third) is the part of the systematic uncertainty that is 
uncorrelated (correlated) between bins.}
\begin{center}
\scalebox{1.0}{
\begin{tabular}{@{}lr@{}c@{}r@{}c@{}r@{}c@{}rrr@{}c@{}r@{}c@{}r@{}c@{}r@{}}
\midrule
 \multicolumn{1}{c}{$y$} & \multicolumn{7}{c}{Prompt $J/\psi$} & 
& \multicolumn{7}{c}{\fromb} \\ \midrule\ $2.0 - 2.5 \ \ $ & $5140 \tpm 26 \tpm 49 \tpm 368$ & & $717 \tpm 6 \tpm 12 \tpm 51$ \\
\ $2.5 - 3.0 \ \ $ & $5066 \tpm 14 \tpm 30 \tpm 363$ & & $666 \tpm 3 \tpm 9 \tpm 48$ \\
\ $3.0 - 3.5 \ \ $ & $4573 \tpm 11 \tpm 20 \tpm 328$ & & $538 \tpm 3 \tpm 8 \tpm 39$ \\
\ $3.5 - 4.0 \ \ $ & $3940 \tpm 11 \tpm 24 \tpm 282$ & & $388 \tpm 2 \tpm 4 \tpm 28$ \\
\ $4.0 - 4.5 \ \ $ & $3153 \tpm 12 \tpm 16 \tpm 226$ & & $240 \tpm 2 \tpm 3 \tpm 17$ \\

\bottomrule 
\end{tabular}
}
\end{center}
\label{tab:rapidity} 
\end{table}
\renewcommand{\arraystretch}{1.}

\begin{figure}[!tb]
\centering
\includegraphics[width=0.48\textwidth]{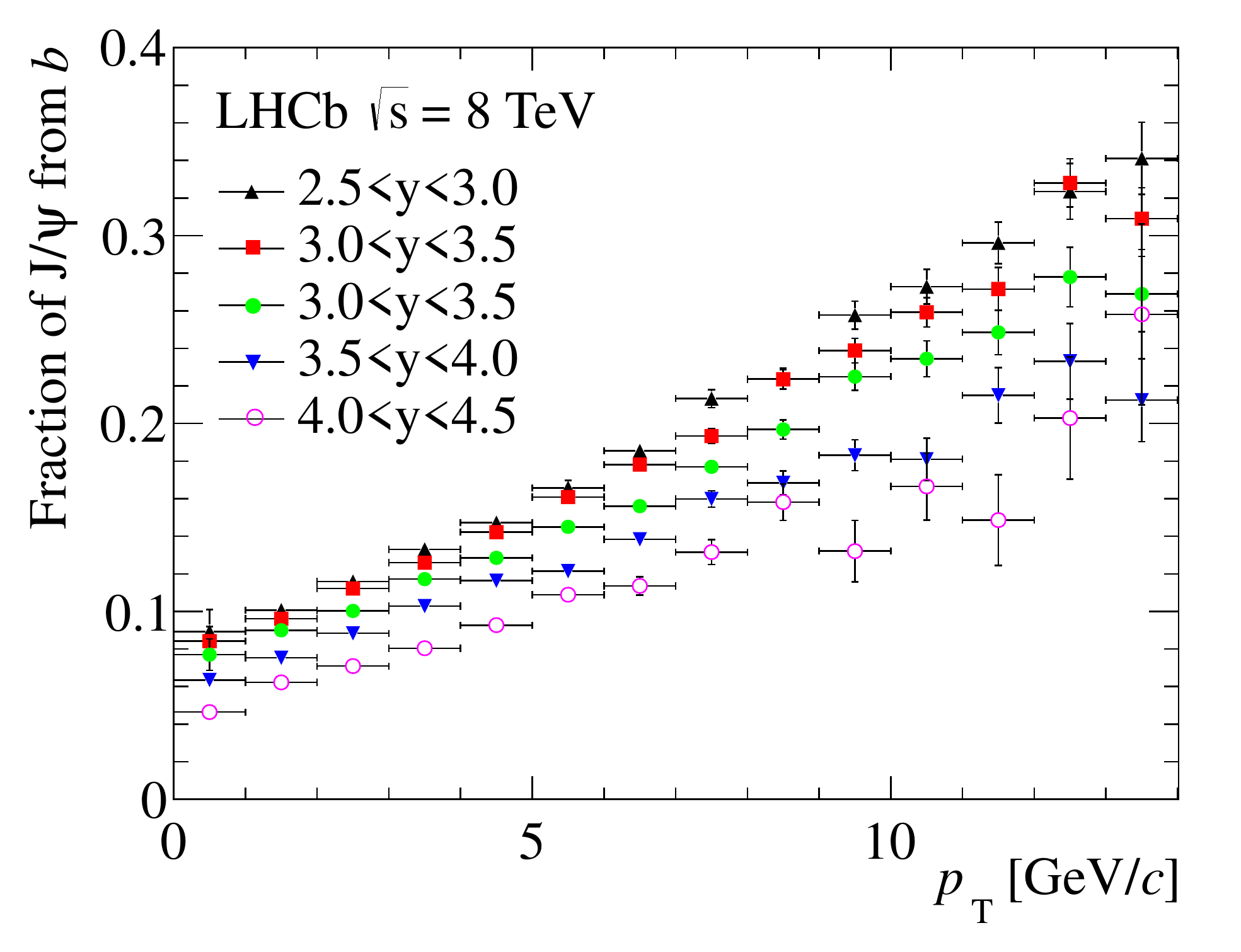}
\caption{\small Fraction of \fromb\ as a function of \pt, in bins of $y$. } \label{fig:bfraction}
\end{figure}

Table~\ref{tab:bfraction} and Fig.~\ref{fig:bfraction} give the values of the fraction of \fromb\ in the different bins of \pt\ and $y$. The fraction of \jpsi\ mesons from $b$-hadron decays increases as a function of \pt, and, at constant \pt,  decreases with increasing $y$, 
as seen in the study at $\sqrt{s} = 7\,\tev$~\cite{LHCb-PAPER-2011-003}.

\section{Results on {\boldmath \ups} meson production} 
\label{upsresults}
\begin{figure}[!H!]
\begin{center}
\includegraphics[width=0.6\textwidth]{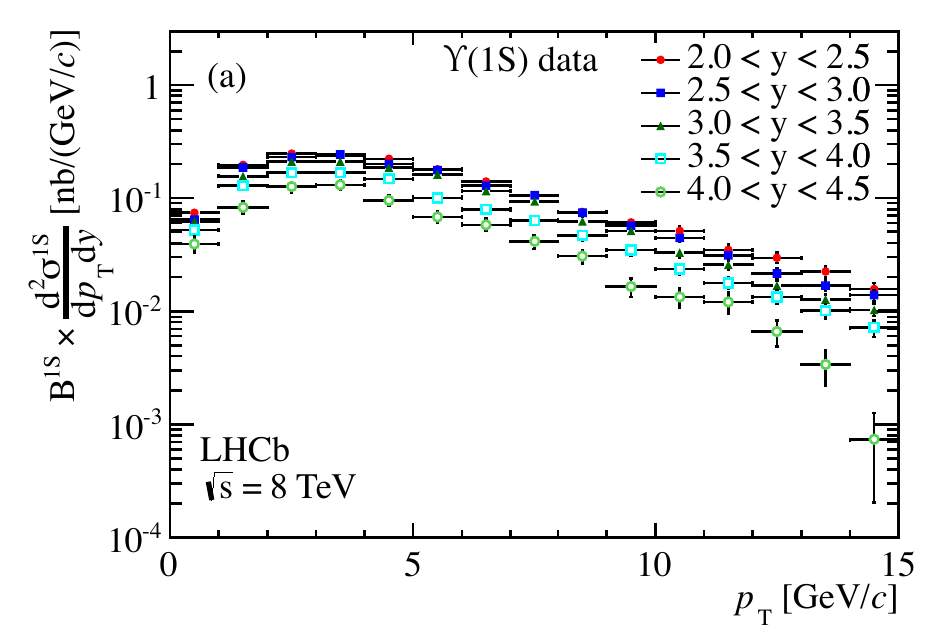}\\
\includegraphics[width=0.6\textwidth]{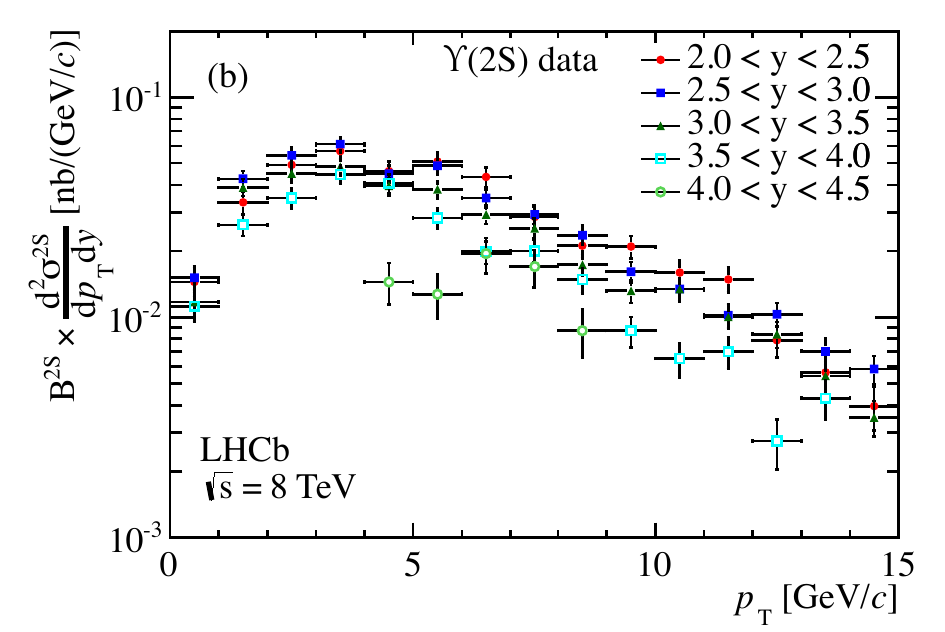} \\ 
\includegraphics[width=0.6\textwidth]{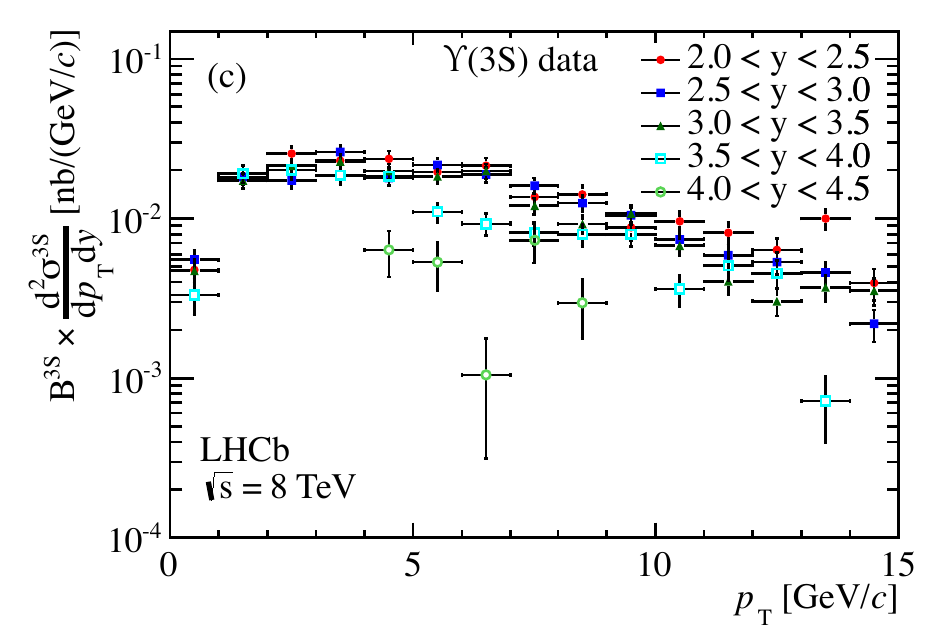} 
 \end{center}
 \caption{\small Double-differential cross-sections times dimuon branching fractions as a function of \pt\ in bins of $y$ 
for (a) the $\ones$, (b) $\twos$ and (c) $\threes$ mesons. \label{fig:xsecY}}
\end{figure}

\begin{figure}[t]
\bce
  \includegraphics[width=0.6\textwidth]{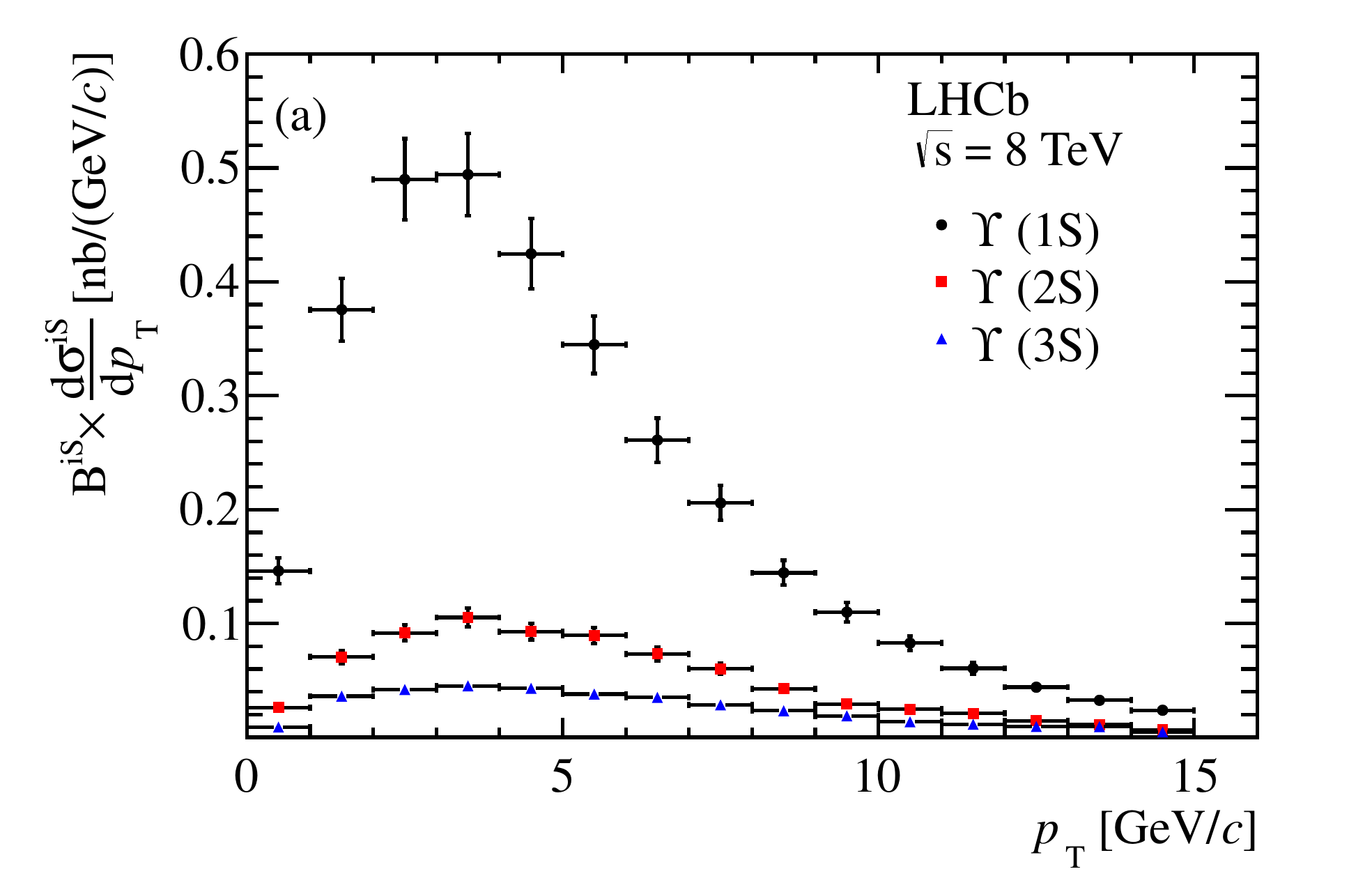}
 \includegraphics[width=0.6\textwidth]{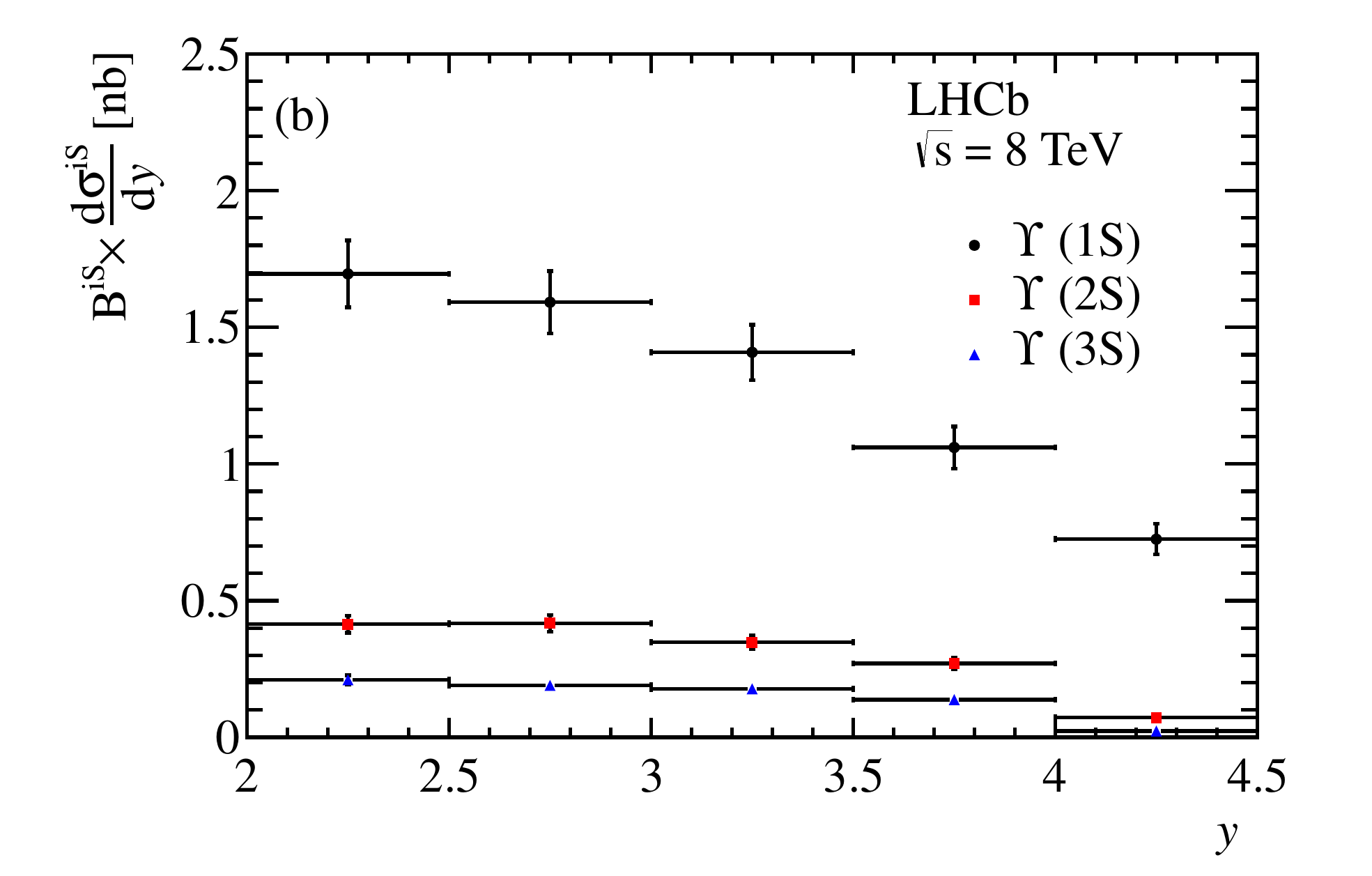}
 \caption{\small Differential production cross-sections for  $\ones$, $\twos$ and
   $\threes$ mesons times dimuon branching fraction (a) as a function of
    $\pt$ integrated over $y$, and (b) as a function of $y$
   integrated over $\pt$. }\label{fig::allups} \ece
\end{figure}

\begin{figure}[b]
\bce
 \includegraphics[width=0.6\textwidth]{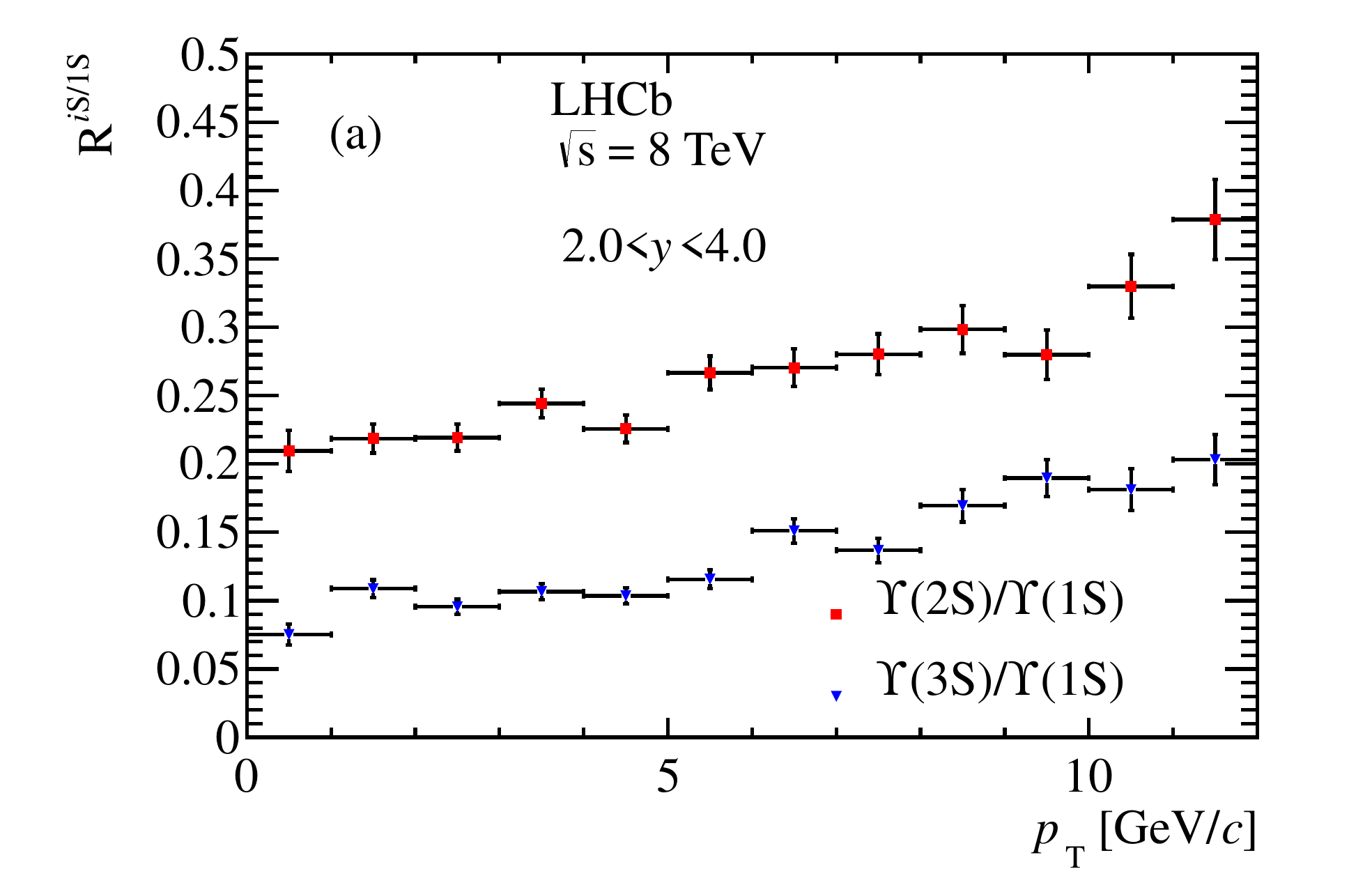}
 \includegraphics[width=0.6\textwidth]{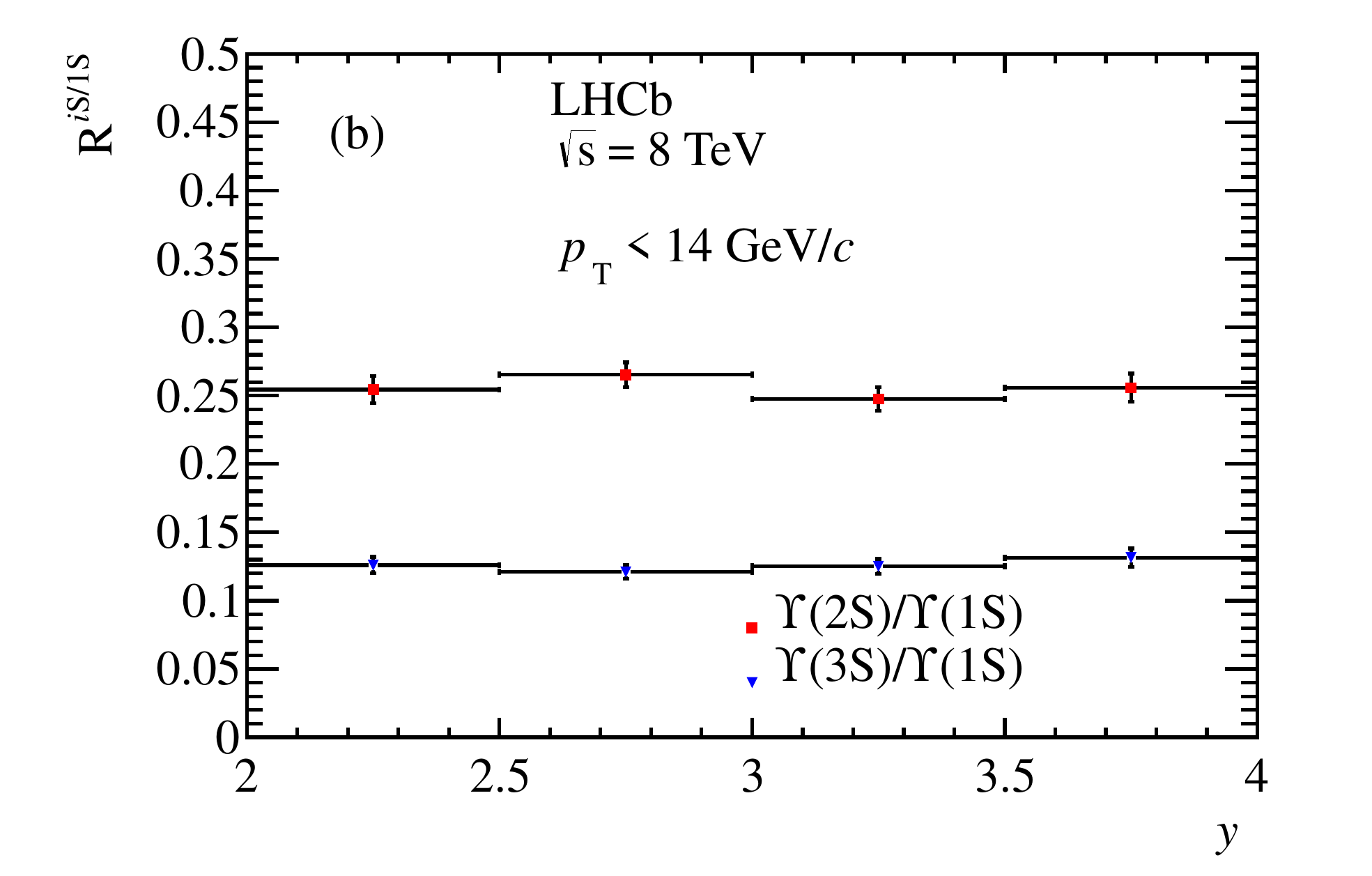}
 \caption{\small {\it } Ratio of the $\twos$ to $\ones$ and $\threes$
   to $\ones$ cross-sections times dimuon branching fractions (a) as a
   function of $\pt$ integrated over $y$, and (b) as a function of
   $y$ integrated over $\pt$.  }\label{fig::myratios} \ece
\end{figure}

The double-differential production cross-sections times the dimuon branching fractions for the $\ups$ mesons  in bins of \pt\ and  $y$ are given in Tables~\ref{tab::xs_ones}, \ref{tab::xs_twos}, and \ref{tab::xs_threes}, with the assumption of  no polarisation.
The double-differential cross-sections are displayed in Fig.~\ref{fig:xsecY}. 
The integrated cross-sections times dimuon branching fractions \nobreak{$B^{iS} = B(\varUpsilon(iS)\to\mu\mu)$}, with $i=1,2,3$, 
 in the range $\pt <15\,\gevc$ and $2.0<y<4.5$ 
are measured to be
\begin{eqnarray*}
\sigma(pp\to \ones X)\times B^{\rm1S} \,& = &\, 3.241\pm 0.018\pm 0.231\,{\rm nb}, \\
\sigma(pp\to \twos X)\times B^{\rm2S} \, &= &\, 0.761\pm 0.008\pm 0.055\,{\rm nb}, \\
\sigma(pp\to \threes X)\times B^{\rm3S} \,& = &\, 0.369\pm 0.005\pm 0.027\,{\rm nb},
\end{eqnarray*}
where the first uncertainty is statistical and the second systematic.
The cross-section times dimuon branching fractions for the three
$\ups$ states are compared in Fig.~\ref{fig::allups} as a function of
 \pt\ and $y$.
These results are used to evaluate the ratios $R^{iS/1S}$ of the $\twos$ to $\ones$ and $\threes$ to $\ones$
cross-sections times dimuon branching fractions.  Most of the uncertainties cancel in the ratio, except those due to the
size of the data sample, to the model dependence and to the choice of the fit function. The ratios $R^{iS/1S}$ as a 
function 
\clearpage
\noindent
of \pt\ and $y$ are given in Tables~\ref{tab::ratiospt} and~\ref{tab::ratiosy}, respectively, and
shown in Fig.~\ref{fig::myratios}, with the assumption of no
polarisation.  For this measurement the $\pt$ range has been restricted to $\pt<14$\,\gevc 
and the $y$ range to $2.0<y<4.0$ to ensure 
enough counts for the three $\ups$ states in all bins.
The ratios are constant as a function of $y$ and increase as a function 
of $\pt$, in agreement with previous observations by LHCb~\cite{LHCB-PAPER-2011-036}
and as reported by ATLAS~\cite{AtlasUpsilon} and CMS~\cite{CmsUpsilon} at $\sqrt{s}=7\,\tev$.

\clearpage
\section{Comparison with theoretical models}
\begin{figure}[b!]
\begin{center}
\includegraphics[width=0.6\textwidth]{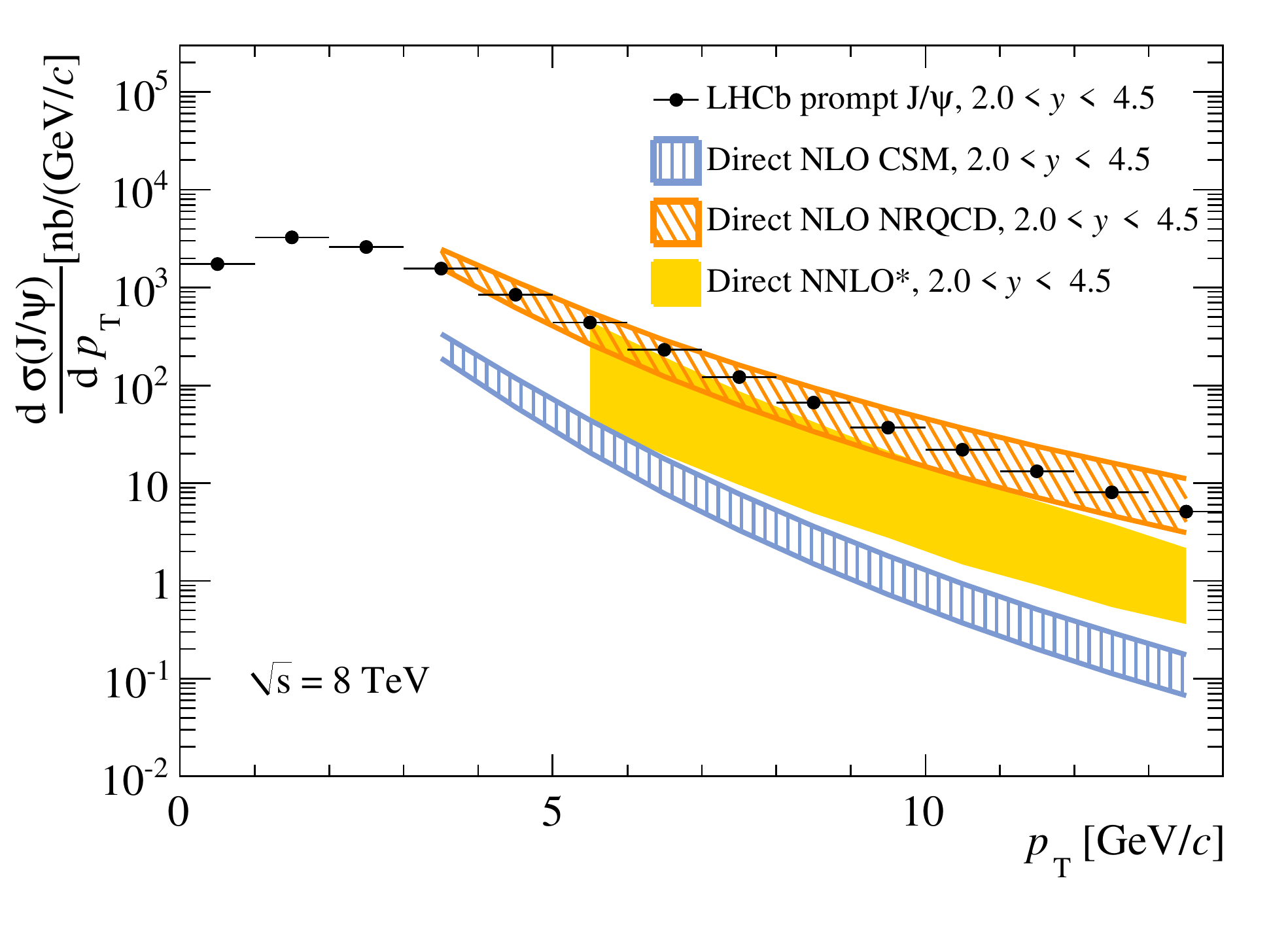}
 \end{center}
\vspace*{-0.9cm}
 \caption{\small Comparison of the differential cross-section for
   the production of prompt \jpsi\ meson (under the assumption of zero polarisation) as a function of
   $\pt$ \label{fig:nlo} with direct production in an NLO NRQCD model~\cite{Butenschoen:2011yh, Butenschoen:2010rq}
   (orange diagonal shading), an NNLO* CSM~\cite{lansberg:2009} (solid yellow) and an
   NLO CSM~\cite{Campbell:2007ws}  (blue vertical shading). The points
 show the measurements reported in this analysis.
}
\end{figure}

The measured differential cross-sections for the production of prompt \jpsi\ mesons as a function of $\pt$ are compared in Fig.~\ref{fig:nlo}  to three theoretical models that assume no polarisation. The considered models are
\begin{itemize}
\renewcommand{\labelitemi}{--}
\item an NRQCD model at next-to-leading order (NLO). The colour-octet matrix elements in this case are determined from 
a global fit to HERA, Tevatron and LHC data~\cite{Butenschoen:2011yh, Butenschoen:2010rq};
\item  an NNLO* CSM~\cite{artoisenet:2008,lansberg:2009}; the notation NNLO* indicates that the calculation  at next-to-next leading order is not complete and neglects part of the logarithmic terms;
\item an NLO CSM~\cite{Campbell:2007ws} with the input parameters
  related to the choice of scale and charm quark mass given in Ref.~\cite{Butenschoen:2011yh}. 
\end{itemize}
In these comparisons it should be noted that the predictions are for  direct \jpsi meson production,
whereas the experimental measurements include feed-down from higher charmonium states. In particular, the contribution 
from \jpsi mesons produced in radiative $\chic$ decays in the considered
fiducial range was measured to be at the level of $20\%$ at $\sqrt{s}= 7\,\tev$~\cite{LHCb:2012af}. Allowing for this contribution, as was seen in the
previous studies, both the NNLO* CSM and the NLO NRQCD models provide
reasonable descriptions of the experimental data. In contrast, the CSM at NLO
underestimates the cross-section by an order of magnitude.

\begin{figure}[tb!]
\begin{center}
\includegraphics[width=0.55\textwidth]{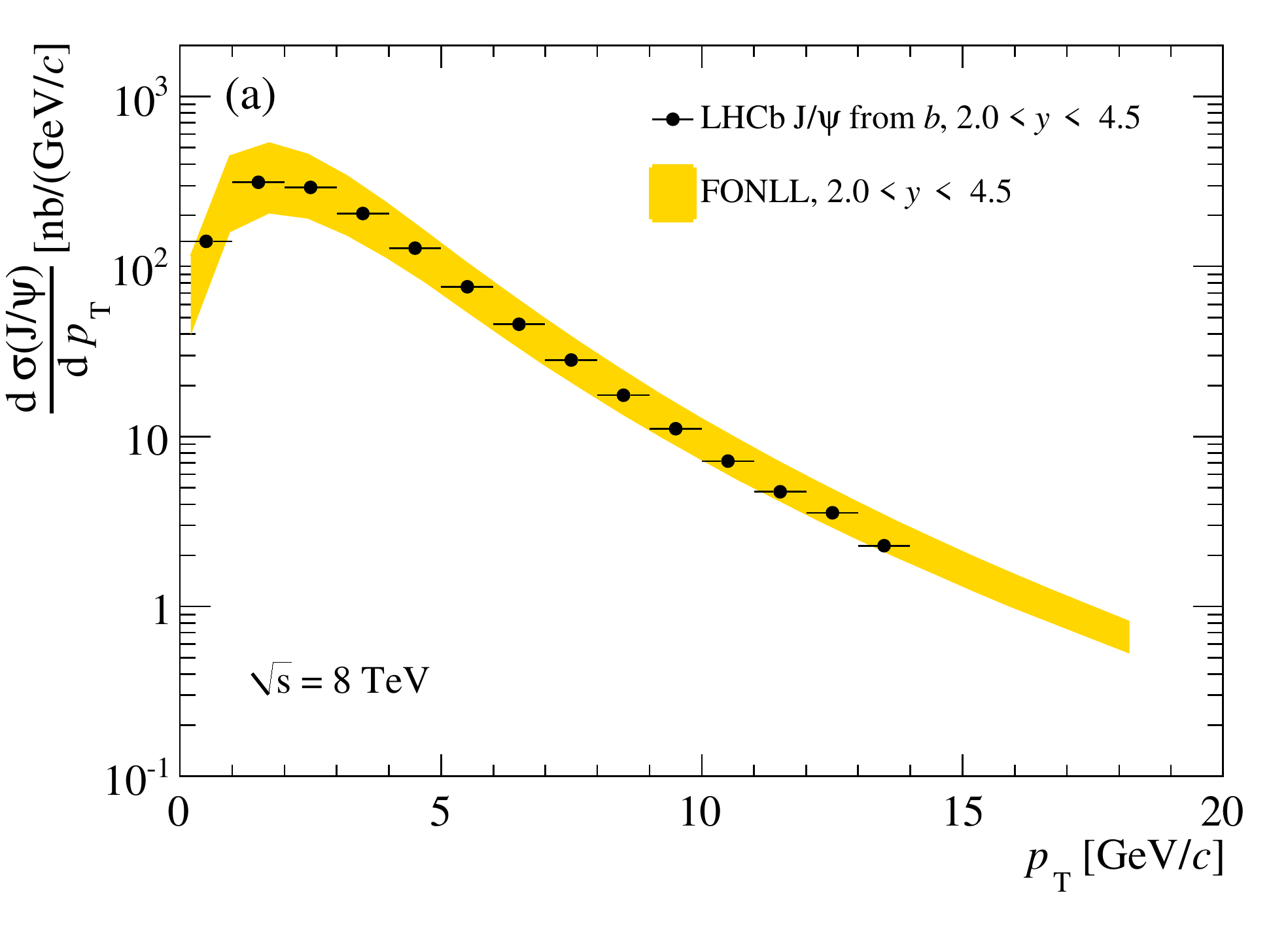}
\includegraphics[width=0.55\textwidth]{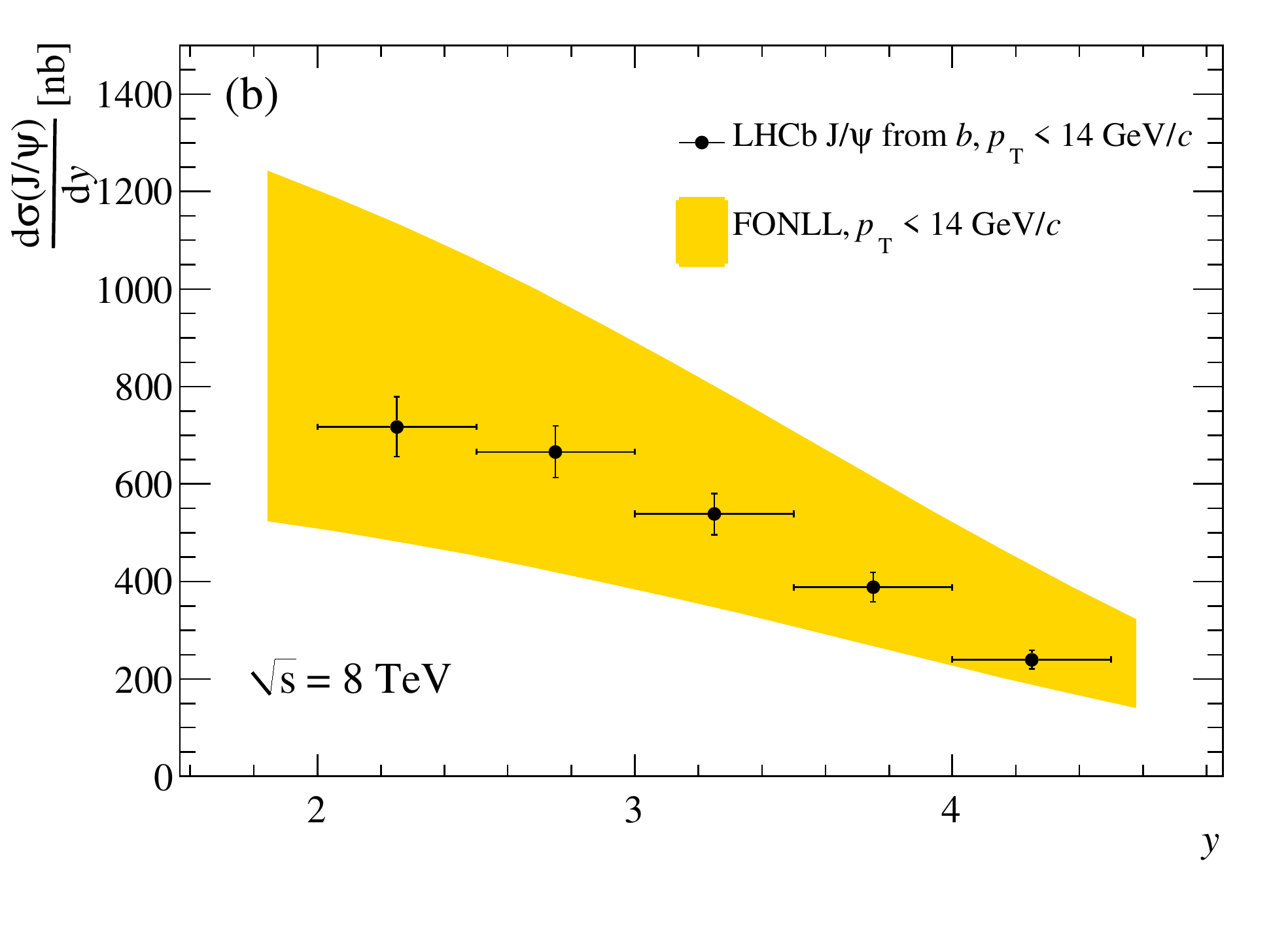}
\vspace*{-0.5cm}
\caption{\small Differential production cross-section for \fromb\ (a) as a function of \pt in the
  fiducial range $2.0<y<4.5$, and (b) as a function of $y$ in the fiducial range $\pt<14\,\gevc$. 
The FONLL prediction~\cite{FONLL,Cacciari} is shown in yellow. The points
 show the measurements reported in this analysis. }
\label{fig:fonlldiff}
\end{center}
\end{figure}

\begin{figure}[t!]
\begin{center}
\includegraphics[width=0.55\textwidth]{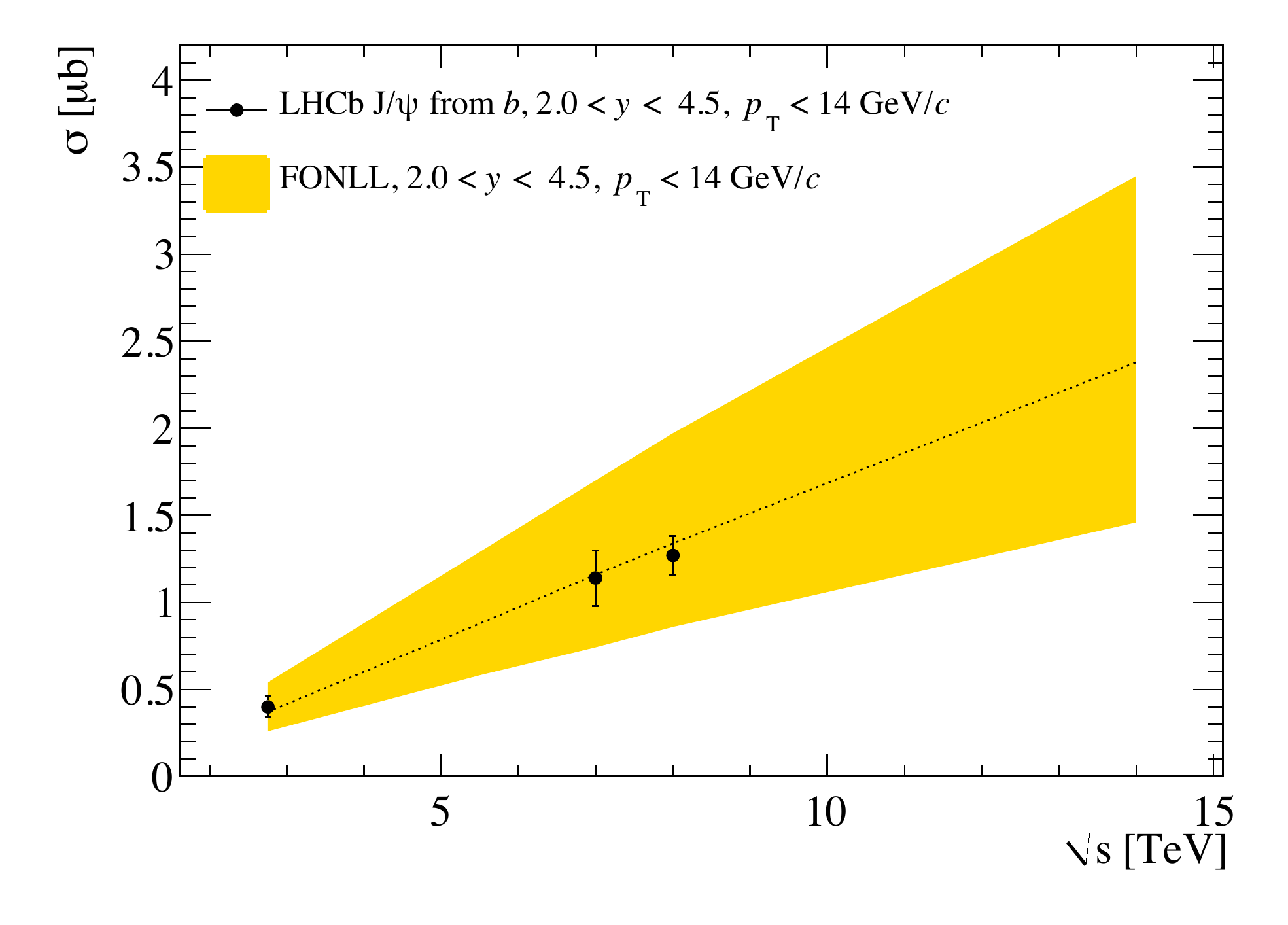}
\caption{\small Predictions based on the FONLL formalism~\cite{FONLL,Cacciari} for the production cross-section for \fromb\ in the fiducial range $0<\pt<14\,\gevc$ and $2.0<y<4.5$ (yellow band). The uncertainty includes contributions from the renormalisation scale, quark masses and the choice of PDF set.  The black dotted line shows the central value of the prediction. The points show the LHCb measurements at $\sqrt{s}=2.76$~\cite{LHCb-PAPER-2012-039}, 7~\cite{LHCb-PAPER-2011-003},  and 8$\,\tev$. }
\label{fig:fonlls}
\end{center}
\end{figure}

The results for the production of \fromb\ can be
compared to calculations based on the FONLL formalism~\cite{FONLL,Cacciari}.  This model predicts
the $b$-quark production cross-section, and includes the fragmentation of the $b$-quark into $b$-hadrons and their 
decay into \jpsi\ mesons. 
In Fig.~\ref{fig:fonlldiff} the data for the differential production cross-section  as a function of \pt\ and $y$  at $\sqrt{s} = 8\,\tev$  are compared to the FONLL predictions. Good agreement is observed.
The prediction for the total cross-section in the
fiducial range of this measurement is $1.34^{+0.63}_{-0.49}\,\upmu{\rm
  b}$, in good agreement with the result presented here. In
Fig.~\ref{fig:fonlls} the measurements of the cross-section for \fromb\
at  $\sqrt{s}=2.76$~\cite{LHCb-PAPER-2012-039}, 7~\cite{LHCb-PAPER-2011-003}, and $8\,\tev$ are compared to the
FONLL prediction. The behaviour as a function of the centre-of-mass energy is in
excellent agreement with the prediction.  This gives confidence
that this model can produce reliable predictions for the $b$-hadron
cross-section at the higher energies expected at the LHC.  

\begin{figure}[bt!]
\bce
  \includegraphics[width=0.55\textwidth, height=6cm]{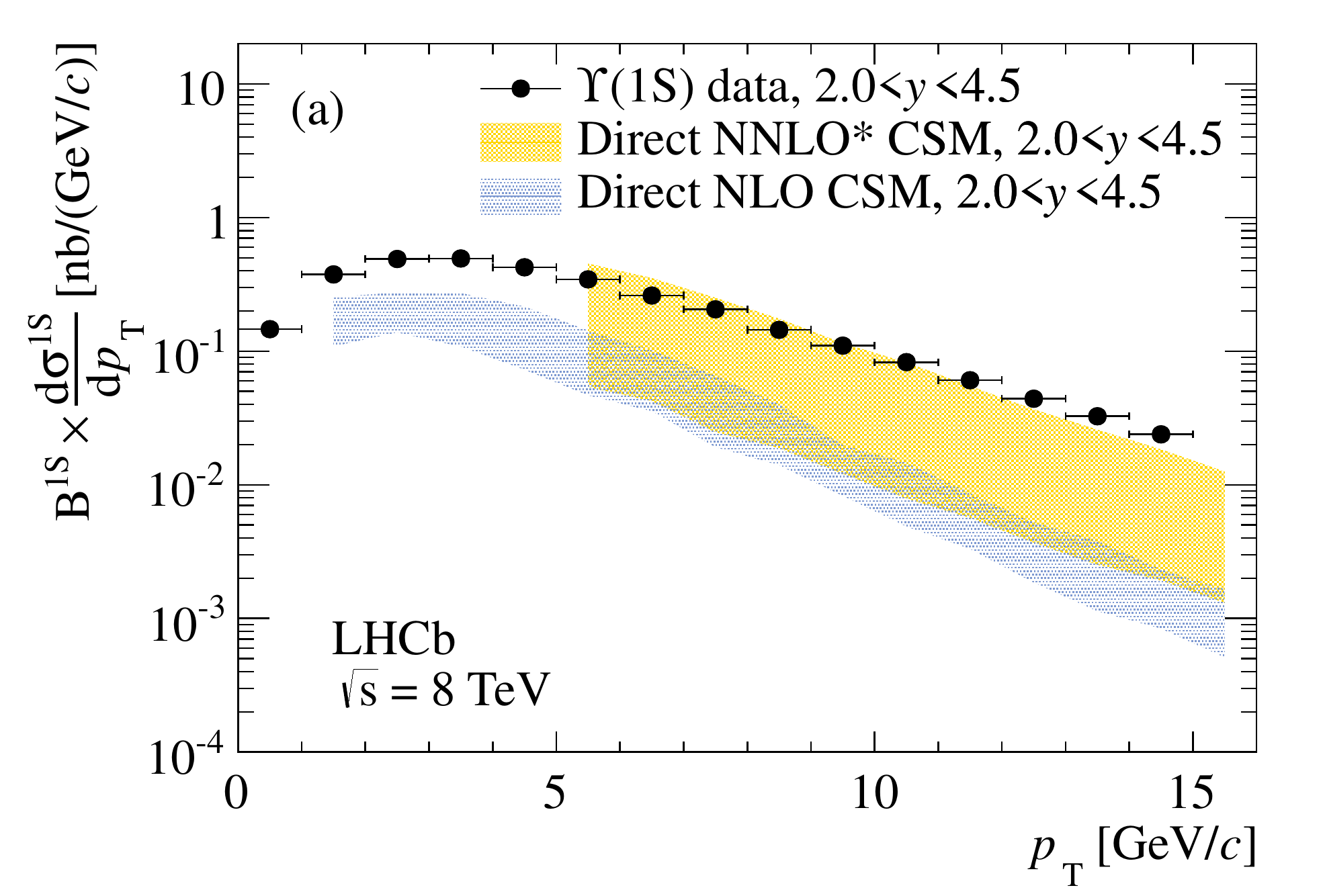}\\
  \includegraphics[width=0.55\textwidth, height=6cm]{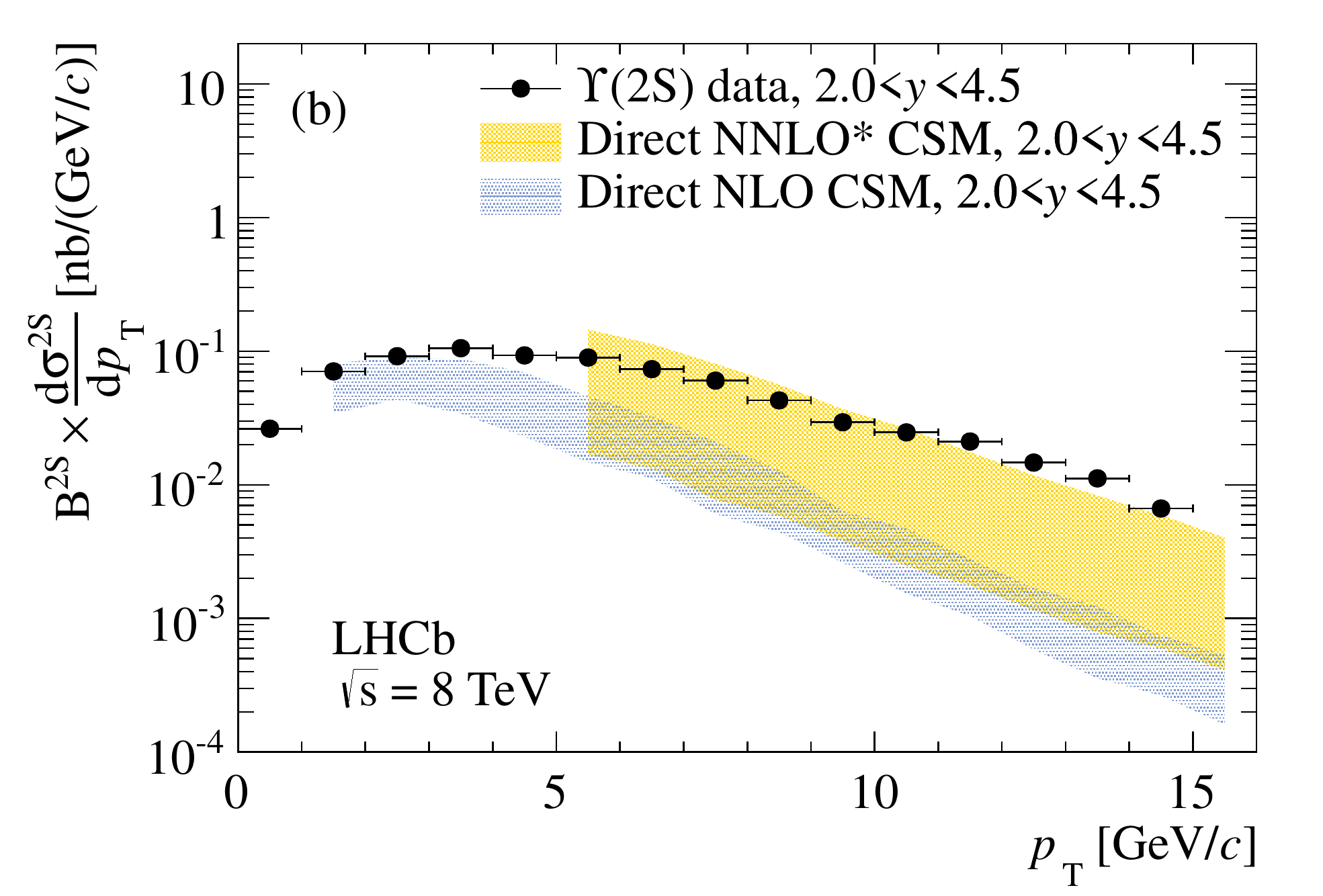}\\
  \includegraphics[width=0.55\textwidth, height=6cm]{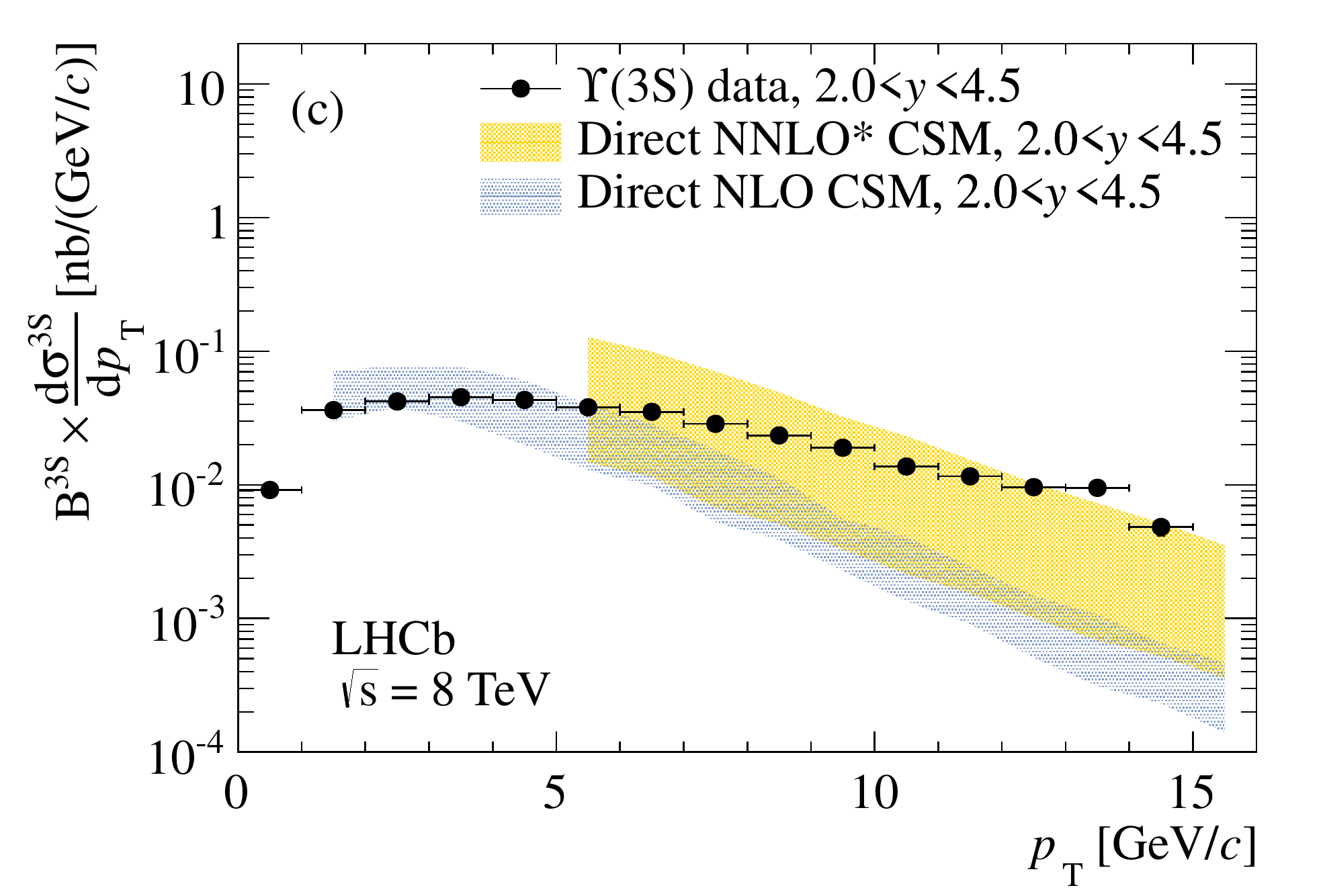}
 \caption{\small Comparison of the differential production cross-sections times dimuon branching fractions for (a) $\ones$, (b) $\twos$ 
  and (c) $\threes$ mesons  as a function of
   $\pt$   with direct production in an NNLO$^*$ CSM~\cite{artoisenet:2008} (solid yellow) and an
   NLO CSM~\cite{Campbell:2007ws} (blue vertical shading) model. The points
 show the measurements reported in this analysis.}\label{fig::csmnloups} \ece
\end{figure}

In Fig.~\ref{fig::csmnloups} the cross-sections times dimuon branching fractions for the 
three $\ups$ meson states are compared to the CSM NLO~\cite{Campbell:2007ws} and NNLO$^*$
theoretical predictions~\cite{artoisenet:2008}
as a function of  $\pt$. The NNLO* CSM  provides a
reasonable description of the experimental data, particularly for the $\threes$ meson, which is expected to be less affected by feed-down. 
As for the prompt \jpsi meson production, the CSM at NLO underestimates the cross-section by an order of magnitude.

\section{Conclusions}

The differential production cross-sections for  \jpsi\ and  $\ups$ mesons are measured as a function of \pt\ and $y$ in the forward region, $2.0<y<4.5$. The analysis is based on a data sample, corresponding to an
integrated luminosity of $18\,\invpb$ and $51\,\invpb$ for the \jpsi\ and  $\ups$ mesons, respectively, collected in the early part of 2012 at a centre-of-mass energy of 
$\sqrt{s}=8\tev$. The  production cross-sections  of \prompt\ mesons and \fromb\  are individually measured. An estimate of the $b\overline{b}$ total cross-section is also obtained.

The results are compared with several recent theoretical predictions in the LHCb acceptance. The NNLO* CSM and the NLO NRQCD model (for the \jpsi\ meson) provide a
reasonable description of the experimental data on the production of prompt \jpsi\  and \ups\ mesons 
as a function of \pt\ under the assumption of zero polarisation. 
This confirms the progress in the theoretical calculations 
of quarkonium hadroproduction, even if the uncertainties on the predictions are still large.
Theoretical predictions based on FONLL calculations are found to describe well
the measured cross-section for \fromb\ and its dependence on centre-of-mass energy.
\clearpage
\section*{Acknowledgements}

\noindent We thank M.~Cacciari for providing the FONLL
predictions for the $b$ cross-section in the LHCb acceptance,
J.~P.~Lansberg for the NNLO* predictions for prompt \jpsi\ and \ups\ meson 
production, B.~Kniehl and M.~Butensch{\"o}n for the NLO NRQCD predictions for prompt
\jpsi\ meson production. We express our gratitude to our colleagues in the CERN
accelerator departments for the excellent performance of the LHC. We
thank the technical and administrative staff at the LHCb
institutes. We acknowledge support from CERN and from the national
agencies: CAPES, CNPq, FAPERJ and FINEP (Brazil); NSFC (China);
CNRS/IN2P3 and Region Auvergne (France); BMBF, DFG, HGF and MPG
(Germany); SFI (Ireland); INFN (Italy); FOM and NWO (The Netherlands);
SCSR (Poland); ANCS/IFA (Romania); MinES, Rosatom, RFBR and NRC
``Kurchatov Institute'' (Russia); MinECo, XuntaGal and GENCAT (Spain);
SNSF and SER (Switzerland); NAS Ukraine (Ukraine); STFC (United
Kingdom); NSF (USA). We also acknowledge the support received from the
ERC under FP7. The Tier1 computing centres are supported by IN2P3
(France), KIT and BMBF (Germany), INFN (Italy), NWO and SURF (The
Netherlands), PIC (Spain), GridPP (United Kingdom). We are thankful
for the computing resources put at our disposal by Yandex LLC
(Russia), as well as to the communities behind the multiple open
source software packages that we depend on.

%
\clearpage
\renewcommand{\arraystretch}{1.35}
\begin{table}[!ht]
\begin{center}
\caption{\small \label{promptresult}Double-differential cross-section $\frac{{\rm d}^2\sigma}{{\rm d}p_{\rm T}{\rm d}y}$ in nb/(\gevc) for prompt \jpsi\ meson production in 
bins of of \pt\ and $y$, with the assumption of no polarisation. The first error is statistical, the second  
is the component of the systematic uncertainty that is uncorrelated between bins and the third is the correlated 
component.}
\scalebox{0.8}{
\begin{tabular}{@{}r@{}c@{}rrr@{}c@{}r@{}c@{}r@{}c@{}rr@{}c@{}r@{}c@{}r@{}c@{}rr@{}c@{}r@{}c@{}r@{}c@{}r@{}}
\midrule
\multicolumn{4}{c}{$\pt\,(\gevc)$} & \multicolumn{7}{c}{$2.0<y<2.5$} & \multicolumn{7}{c}{$2.5<y<3.0$} & 
\multicolumn{7}{c}{$3.0<y<3.5$} \\
\midrule
 $0$&$-$&$1$&& $727.45 \tpm 13.14 \tpm 25.66 \tpm 52.15$ & $772.40 \tpm 6.84 \tpm 11.94 \tpm 55.38$ & $727.58 \tpm 5.57 \tpm 10.31 \tpm 52.16$\\
 $1$&$-$&$2$&& $1463.31 \tpm 16.73 \tpm 19.26 \tpm 104.91$ & $1473.44 \tpm 8.74 \tpm 13.74 \tpm 105.64$ & $1357.44 \tpm 7.01 \tpm 11.10 \tpm 97.32$\\
 $2$&$-$&$3$&& $1237.49 \tpm 12.28 \tpm 22.39 \tpm 88.72$ & $1196.41 \tpm 6.69 \tpm 20.45 \tpm 85.78$ & $1083.40 \tpm 5.33 \tpm 9.91 \tpm 77.67$\\
 $3$&$-$&$4$&& $761.17 \tpm 7.29 \tpm 20.60 \tpm 54.57$ & $738.03 \tpm 4.20 \tpm 8.94 \tpm 52.91$ & $651.06 \tpm 3.45 \tpm 6.62 \tpm 46.68$\\
 $4$&$-$&$5$&& $432.67 \tpm 4.33 \tpm 20.79 \tpm 31.02$ & $408.62 \tpm 2.56 \tpm 6.57 \tpm 29.30$ & $359.82 \tpm 2.20 \tpm 4.36 \tpm 25.80$\\
 $5$&$-$&$6$&& $231.63 \tpm 2.60 \tpm 2.89 \tpm 16.61$ & $217.04 \tpm 1.61 \tpm 5.01 \tpm 15.56$ & $183.07 \tpm 1.38 \tpm 2.95 \tpm 13.12$\\
 $6$&$-$&$7$&& $126.50 \tpm 1.65 \tpm 2.29 \tpm 9.07$ & $116.27 \tpm 1.07 \tpm 4.31 \tpm 8.34$ & $97.25 \tpm 0.94 \tpm 1.76 \tpm 6.97$\\
 $7$&$-$&$8$&& $68.05 \tpm 1.07 \tpm 1.98 \tpm 4.88$ & $63.25 \tpm 0.74 \tpm 1.09 \tpm 4.53$ & $51.21 \tpm 0.64 \tpm 1.23 \tpm 3.67$\\
 $8$&$-$&$9$&& $39.23 \tpm 0.74 \tpm 1.88 \tpm 2.81$ & $34.85 \tpm 0.52 \tpm 0.84 \tpm 2.50$ & $27.34 \tpm 0.45 \tpm 0.98 \tpm 1.96$\\
 $9$&$-$&$10$&& $22.04 \tpm 0.52 \tpm 0.66 \tpm 1.58$ & $19.54 \tpm 0.37 \tpm 0.49 \tpm 1.40$ & $15.08 \tpm 0.32 \tpm 0.70 \tpm 1.08$\\
 $10$&$-$&$11$&& $13.60 \tpm 0.39 \tpm 0.45 \tpm 0.98$ & $11.59 \tpm 0.28 \tpm 0.37 \tpm 0.83$ & $8.88 \tpm 0.25 \tpm 0.15 \tpm 0.64$\\
 $11$&$-$&$12$&& $8.06 \tpm 0.28 \tpm 0.23 \tpm 0.58$ & $7.29 \tpm 0.22 \tpm 0.33 \tpm 0.52$ & $5.03 \tpm 0.18 \tpm 0.11 \tpm 0.36$\\
 $12$&$-$&$13$&& $5.26 \tpm 0.22 \tpm 0.20 \tpm 0.38$ & $4.31 \tpm 0.16 \tpm 0.06 \tpm 0.31$ & $3.30 \tpm 0.14 \tpm 0.10 \tpm 0.24$\\
 $13$&$-$&$14$&& $3.30 \tpm 0.17 \tpm 0.21 \tpm 0.24$ & $2.94 \tpm 0.14 \tpm 0.07 \tpm 0.21$ & $2.09 \tpm 0.11 \tpm 0.09 \tpm 0.15$\\
\hline
&&&& \multicolumn{7}{c}{$3.5<y<4.0$} & \multicolumn{7}{c}{$4.0<y<4.5$}\\ 
\hline
 $0$&$-$&$1$&& $670.50 \tpm 5.34 \tpm 6.78 \tpm 48.07$ & $579.84 \tpm 6.08 \tpm 7.65 \tpm 41.57$ &\\
 $1$&$-$&$2$&& $1213.97 \tpm 6.52 \tpm 20.90 \tpm 87.04$ & $1003.19 \tpm 7.18 \tpm 10.37 \tpm 71.92$ &\\
 $2$&$-$&$3$&& $926.23 \tpm 4.88 \tpm 7.50 \tpm 66.41$ & $753.31 \tpm 5.75 \tpm 6.80 \tpm 54.01$ &\\
 $3$&$-$&$4$&& $542.09 \tpm 3.19 \tpm 5.56 \tpm 38.87$ & $418.29 \tpm 3.89 \tpm 4.61 \tpm 29.99$ &\\
 $4$&$-$&$5$&& $284.82 \tpm 2.00 \tpm 3.44 \tpm 20.42$ & $204.28 \tpm 2.31 \tpm 2.68 \tpm 14.65$ &\\
 $5$&$-$&$6$&& $146.41 \tpm 1.29 \tpm 2.20 \tpm 10.50$ & $96.52 \tpm 1.39 \tpm 2.61 \tpm 6.92$ &\\
 $6$&$-$&$7$&& $73.68 \tpm 0.84 \tpm 1.33 \tpm 5.28$ & $48.40 \tpm 0.90 \tpm 0.92 \tpm 3.47$ &\\
 $7$&$-$&$8$&& $37.39 \tpm 0.56 \tpm 0.82 \tpm 2.68$ & $23.95 \tpm 0.59 \tpm 0.60 \tpm 1.72$ &\\
 $8$&$-$&$9$&& $20.05 \tpm 0.40 \tpm 0.52 \tpm 1.44$ & $11.83 \tpm 0.39 \tpm 0.34 \tpm 0.85$ &\\
 $9$&$-$&$10$&& $11.04 \tpm 0.29 \tpm 0.38 \tpm 0.79$ & $6.64 \tpm 0.27 \tpm 0.24 \tpm 0.48$ &\\
 $10$&$-$&$11$&& $6.24 \tpm 0.20 \tpm 0.28 \tpm 0.45$ & $3.40 \tpm 0.18 \tpm 0.15 \tpm 0.24$ &\\
 $11$&$-$&$12$&& $3.85 \tpm 0.18 \tpm 0.16 \tpm 0.28$ & $2.05 \tpm 0.14 \tpm 0.10 \tpm 0.15$ &\\
 $12$&$-$&$13$&& $2.23 \tpm 0.13 \tpm 0.05 \tpm 0.16$ & $1.04 \tpm 0.09 \tpm 0.03 \tpm 0.07$ &\\
 $13$&$-$&$14$&& $1.49 \tpm 0.10 \tpm 0.04 \tpm 0.11$ & $0.45 \tpm 0.06 \tpm 0.02 \tpm 0.03$ &\\

\bottomrule
\end{tabular}
}
\end{center}
\end{table}
\renewcommand{\arraystretch}{1.35}
\begin{table}[!ht]
\begin{center}
\caption{\small \label{bresult}Double-differential cross-section $\frac{{\rm d}^2\sigma}{{\rm d}p_{\rm T}{\rm d}y}$ in nb/(\gevc) for the production of   \fromb\ in bins of \pt\ and $y$. The first error is statistical, the second is the component of the systematic uncertainty that is uncorrelated between bins and the third is the correlated component.}
\scalebox{0.8}{
\begin{tabular}{@{}r@{}c@{}rrr@{}c@{}r@{}c@{}r@{}c@{}rr@{}c@{}r@{}c@{}r@{}c@{}rr@{}c@{}r@{}c@{}r@{}c@{}r@{}}
\midrule
\multicolumn{4}{c}{$\pt\,(\gevc)$} & \multicolumn{7}{c}{$2.0<y<2.5$} & \multicolumn{7}{c}{$2.5<y<3.0$} & \multicolumn{7}{c}{$3.0<y<3.5$} \\
\hline
%
%
 $0$&$-$&$1$&& $71.82 \tpm 2.53 \tpm 9.44 \tpm 5.15$ & $71.70 \tpm 1.33 \tpm 6.58 \tpm 5.14$ & $61.63 \tpm 1.10 \tpm 6.63 \tpm 4.42$\\
 $1$&$-$&$2$&& $164.48 \tpm 3.49 \tpm 3.71 \tpm 11.79$ & $157.59 \tpm 1.84 \tpm 3.88 \tpm 11.30$ & $135.22 \tpm 1.46 \tpm 2.56 \tpm 9.69$\\
 $2$&$-$&$3$&& $162.88 \tpm 2.95 \tpm 3.70 \tpm 11.68$ & $152.06 \tpm 1.62 \tpm 3.38 \tpm 10.90$ & $121.63 \tpm 1.27 \tpm 2.08 \tpm 8.72$\\
 $3$&$-$&$4$&& $117.14 \tpm 2.02 \tpm 3.49 \tpm 8.40$ & $106.89 \tpm 1.16 \tpm 1.79 \tpm 7.66$ & $87.13 \tpm 0.95 \tpm 1.50 \tpm 6.25$\\
 $4$&$-$&$5$&& $75.00 \tpm 1.34 \tpm 3.71 \tpm 5.38$ & $68.17 \tpm 0.81 \tpm 1.29 \tpm 4.89$ & $53.63 \tpm 0.67 \tpm 0.87 \tpm 3.84$\\
 $5$&$-$&$6$&& $46.32 \tpm 0.91 \tpm 0.97 \tpm 3.32$ & $41.94 \tpm 0.57 \tpm 1.07 \tpm 3.01$ & $31.46 \tpm 0.48 \tpm 0.64 \tpm 2.26$\\
 $6$&$-$&$7$&& $28.96 \tpm 0.64 \tpm 0.56 \tpm 2.08$ & $25.49 \tpm 0.42 \tpm 0.96 \tpm 1.83$ & $18.30 \tpm 0.35 \tpm 0.37 \tpm 1.31$\\
 $7$&$-$&$8$&& $18.59 \tpm 0.46 \tpm 0.57 \tpm 1.33$ & $15.36 \tpm 0.31 \tpm 0.31 \tpm 1.10$ & $11.25 \tpm 0.26 \tpm 0.27 \tpm 0.81$\\
 $8$&$-$&$9$&& $11.40 \tpm 0.34 \tpm 0.55 \tpm 0.82$ & $10.20 \tpm 0.25 \tpm 0.27 \tpm 0.73$ & $6.88 \tpm 0.20 \tpm 0.25 \tpm 0.49$\\
 $9$&$-$&$10$&& $7.77 \tpm 0.27 \tpm 0.24 \tpm 0.56$ & $6.26 \tpm 0.19 \tpm 0.17 \tpm 0.45$ & $4.54 \tpm 0.16 \tpm 0.21 \tpm 0.33$\\
 $10$&$-$&$11$&& $5.23 \tpm 0.22 \tpm 0.18 \tpm 0.38$ & $4.16 \tpm 0.15 \tpm 0.13 \tpm 0.30$ & $2.83 \tpm 0.13 \tpm 0.06 \tpm 0.20$\\
 $11$&$-$&$12$&& $3.46 \tpm 0.17 \tpm 0.10 \tpm 0.25$ & $2.82 \tpm 0.13 \tpm 0.14 \tpm 0.20$ & $1.75 \tpm 0.10 \tpm 0.04 \tpm 0.13$\\
 $12$&$-$&$13$&& $2.61 \tpm 0.14 \tpm 0.11 \tpm 0.19$ & $2.17 \tpm 0.11 \tpm 0.03 \tpm 0.16$ & $1.35 \tpm 0.09 \tpm 0.05 \tpm 0.10$\\
 $13$&$-$&$14$&& $1.76 \tpm 0.11 \tpm 0.12 \tpm 0.13$ & $1.39 \tpm 0.09 \tpm 0.04 \tpm 0.10$ & $0.85 \tpm 0.07 \tpm 0.04 \tpm 0.06$\\
\hline
&&&& \multicolumn{7}{c}{$3.5<y<4.0$} & \multicolumn{7}{c}{$4.0<y<4.5$}\\ 
\hline
 $0$&$-$&$1$&& $46.51 \tpm 1.04 \tpm 1.11 \tpm 3.33$ & $29.15 \tpm 1.03 \tpm 1.41 \tpm 2.09$ &\\
 $1$&$-$&$2$&& $100.11 \tpm 1.29 \tpm 3.76 \tpm 7.18$ & $67.76 \tpm 1.35 \tpm 2.77 \tpm 4.86$ &\\
 $2$&$-$&$3$&& $90.70 \tpm 1.20 \tpm 1.40 \tpm 6.50$ & $58.35 \tpm 1.22 \tpm 0.72 \tpm 4.18$ &\\
 $3$&$-$&$4$&& $62.80 \tpm 0.89 \tpm 1.38 \tpm 4.50$ & $37.05 \tpm 0.90 \tpm 0.49 \tpm 2.66$ &\\
 $4$&$-$&$5$&& $37.98 \tpm 0.59 \tpm 0.62 \tpm 2.72$ & $21.13 \tpm 0.61 \tpm 0.44 \tpm 1.52$ &\\
 $5$&$-$&$6$&& $20.61 \tpm 0.41 \tpm 0.37 \tpm 1.48$ & $12.04 \tpm 0.42 \tpm 0.34 \tpm 0.86$ &\\
 $6$&$-$&$7$&& $12.09 \tpm 0.30 \tpm 0.22 \tpm 0.87$ & $6.38 \tpm 0.29 \tpm 0.12 \tpm 0.46$ &\\
 $7$&$-$&$8$&& $7.27 \tpm 0.22 \tpm 0.16 \tpm 0.52$ & $3.69 \tpm 0.20 \tpm 0.09 \tpm 0.26$ &\\
 $8$&$-$&$9$&& $4.19 \tpm 0.17 \tpm 0.12 \tpm 0.30$ & $2.27 \tpm 0.15 \tpm 0.07 \tpm 0.16$ &\\
 $9$&$-$&$10$&& $2.56 \tpm 0.13 \tpm 0.09 \tpm 0.18$ & $1.09 \tpm 0.12 \tpm 0.09 \tpm 0.08$ &\\
 $10$&$-$&$11$&& $1.42 \tpm 0.09 \tpm 0.07 \tpm 0.10$ & $0.72 \tpm 0.08 \tpm 0.04 \tpm 0.05$ &\\
 $11$&$-$&$12$&& $1.09 \tpm 0.08 \tpm 0.05 \tpm 0.08$ & $0.37 \tpm 0.05 \tpm 0.04 \tpm 0.03$ &\\
 $12$&$-$&$13$&& $0.73 \tpm 0.07 \tpm 0.02 \tpm 0.05$ & $0.28 \tpm 0.04 \tpm 0.02 \tpm 0.02$ &\\
 $13$&$-$&$14$&& $0.42 \tpm 0.05 \tpm 0.02 \tpm 0.03$ & $0.16 \tpm 0.03 \tpm 0.01 \tpm 0.01$ &\\

\bottomrule
\end{tabular}
}
\end{center}
\end{table}
\renewcommand{\arraystretch}{1.2}
\renewcommand{\arraystretch}{1.35}
\renewcommand{\arraystretch}{1.44}
\begin{table}[!ht]
\begin{center}
\caption{\small \label{tab:bfraction}Fraction of \fromb\ (in \%) in bins of  of \pt\ and $y$. 
The first uncertainty is statistical and the second systematic (uncorrelated between bins).}
\scalebox{0.85}{
\begin{tabular}{@{}r@{}c@{}rrr@{}c@{}r@{}c@{}r@{}rr@{}c@{}r@{}c@{}r@{}rr@{}c@{}r@{}c@{}r@{}r@{}}
\midrule
\multicolumn{4}{c}{$\pt\,(\gevc)$} & \multicolumn{6}{c}{$2.0<y<2.5$} & \multicolumn{6}{c}{$2.5<y<3.0$} & \multicolumn{6}{c}{$3.0<y<3.5$} \\
\midrule
$0$&$-$&$1$  && $8.9\tpm 0.3\tpm 1.1$  && $8.4\tpm 0.1\tpm 0.8$ & & $7.7\tpm 0.1\tpm 0.8$& \\
$1$&$-$&$2$  && $10.1\tpm 0.2\tpm 0.2$&& $9.6\tpm 0.1\tpm 0.2$ & & $9.0\tpm 0.1\tpm 0.2$& \\
$2$&$-$&$3$  && $11.6\tpm 0.2\tpm 0.2$&& $11.2\tpm 0.1\tpm 0.2$&& $10.0\tpm 0.1\tpm 0.1$& \\
$3$&$-$&$4$  && $13.3\tpm 0.2\tpm 0.2$& & $12.6\tpm 0.1\tpm 0.1$& & $11.7\tpm 0.1\tpm 0.2$& \\
$4$&$-$&$5$  && $14.7\tpm 0.2\tpm 0.2$& & $14.2\tpm 0.1\tpm 0.1$& & $12.9\tpm 0.1\tpm 0.1$& \\
$5$&$-$&$6$  && $16.6\tpm 0.3\tpm 0.3$& & $16.1\tpm 0.2\tpm 0.2$& & $14.5\tpm 0.2\tpm 0.2$& \\
$6$&$-$&$7$  && $18.5\tpm 0.3\tpm 0.1$& & $17.8\tpm 0.3\tpm 0.1$& & $15.6\tpm 0.3\tpm 0.2$& \\
$7$&$-$&$8$  && $21.3\tpm 0.4\tpm 0.2$& & $19.3\tpm 0.3\tpm 0.2$& & $17.7\tpm 0.3\tpm 0.1$& \\
$8$&$-$&$9$  && $22.4\tpm 0.5\tpm 0.1$& & $22.4\tpm 0.5\tpm 0.2$& & $19.7\tpm 0.5\tpm 0.1$& \\
$9$&$-$&$10$  && $25.8\tpm 0.7\tpm 0.2$& & $23.9\tpm 0.6\tpm 0.2$& & $22.5\tpm 0.7\tpm 0.2$& \\
$10$&$-$&$11$  && $27.3\tpm 0.9\tpm 0.2$& & $25.9\tpm 0.8\tpm 0.1$& & $23.4\tpm 0.9\tpm 0.3$& \\
$11$&$-$&$12$  && $29.6\tpm 1.1\tpm 0.2$& & $27.2\tpm 1.0\tpm 0.5$& & $24.9\tpm 1.2\tpm 0.1$&\\
$12$&$-$&$13$  && $32.3\tpm 1.3\tpm 0.6$& & $32.8\tpm 1.3\tpm 0.2$& & $27.8\tpm 1.5\tpm 0.5$& \\
$13$&$-$&$14$  && $34.1\tpm 1.7\tpm 1.0$& & $30.9\tpm 1.6\tpm 0.4$& & $26.9\tpm 2.0\tpm 0.4$&\\
\midrule
& & & & \multicolumn{6}{c}{$3.5<y<4.0$} & \multicolumn{6}{c}{$4.0<y<4.5$} \\ \midrule
$0$&$-$&$1$  && $6.4\tpm 0.1\tpm 0.1$& & $4.6\tpm 0.2\tpm 0.2$& \\
$1$&$-$&$2$  && $7.5\tpm 0.1\tpm 0.3$& & $6.2\tpm 0.1\tpm 0.2$& \\
$2$&$-$&$3$  && $8.8\tpm 0.1\tpm 0.1$& & $7.1\tpm 0.1\tpm 0.1$& \\
$3$&$-$&$4$  && $10.3\tpm 0.1\tpm 0.2$& & $8.0\tpm 0.2\tpm 0.1$& \\
$4$&$-$&$5$  && $11.6\tpm 0.2\tpm 0.1$& & $9.3\tpm 0.2\tpm 0.1$& \\
$5$&$-$&$6$  && $12.1\tpm 0.2\tpm 0.1$& & $10.9\tpm 0.4\tpm 0.1$& \\
$6$&$-$&$7$  && $13.8\tpm 0.3\tpm 0.1$& & $11.4\tpm 0.5\tpm 0.0$& \\
$7$&$-$&$8$  && $16.0\tpm 0.4\tpm 0.1$& & $13.2\tpm 0.7\tpm 0.1$& \\
$8$&$-$&$9$  && $16.8\tpm 0.6\tpm 0.2$& & $15.8\tpm 0.9\tpm 0.2$& \\
$9$&$-$&$10$  && $18.3\tpm 0.8\tpm 0.1$& & $13.2\tpm 1.3\tpm 0.9$& \\
$10$&$-$&$11$  && $18.1\tpm 1.1\tpm 0.3$& & $16.7\tpm 1.7\tpm 0.6$& \\
$11$&$-$&$12$  && $21.5\tpm 1.5\tpm 0.2$& & $14.9\tpm 2.0\tpm 1.3$& \\
$12$&$-$&$13$  && $23.3\tpm 2.0\tpm 0.4$& & $20.3\tpm 3.0\tpm 1.1$& \\
$13$&$-$&$14$  && $21.2\tpm 2.1\tpm 0.6$& & $25.8\tpm 4.4\tpm 1.9$&\\
\bottomrule
\end{tabular}
}
\end{center}
\end{table}
\renewcommand{\arraystretch}{1.2}
\label{upsilontables}
\renewcommand{\arraystretch}{1.35}
\begin{table}[!ht]
\begin{center}
\caption{\small \label{tab::xs_ones} Double-differential production cross-sections $\frac{{\rm d}^2\sigma}{{\rm d}p_{\rm T}{\rm d}y}\times \mathcal{B}^{1S}$ in pb/(\gevc) for the \ones\ meson in 
bins of transverse momentum and rapidity, assuming no polarisation. 
The first error is statistical, the second is the component of the systematic uncertainty that is uncorrelated between bins and the third is the correlated component.}
\scalebox{0.8}{
\begin{tabular}{cccc}
\\
\midrule
$\pt\,(\gevc)$ 
& $2.0<y<2.5$ 
& $2.5<y<3.0$ 
& $3.0<y<3.5$ \\
\midrule
 0$-$1  &    73.9 $\pm$     4.2 $\pm$     1.8 $\pm$     5.3  &    64.7 $\pm$     2.6 $\pm$     1.3 $\pm$     4.6  &    62.1 $\pm$     2.6 $\pm$     0.7 $\pm$     4.4 \\
 1$-$2  &   197.7 $\pm$     6.8 $\pm$     2.7 $\pm$    14.1  &   185.6 $\pm$     4.5 $\pm$     1.9 $\pm$    13.2  &   157.2 $\pm$     4.2 $\pm$     1.7 $\pm$    11.2 \\
 2$-$3  &   248.1 $\pm$     7.5 $\pm$     2.7 $\pm$    17.6  &   228.5 $\pm$     5.0 $\pm$     2.1 $\pm$    16.2  &   210.6 $\pm$     4.9 $\pm$     1.6 $\pm$    15.0 \\
 3$-$4  &   237.9 $\pm$     7.2 $\pm$     4.1 $\pm$    16.9  &   243.5 $\pm$     5.1 $\pm$     1.5 $\pm$    17.3  &   210.3 $\pm$     4.9 $\pm$     2.2 $\pm$    15.0 \\
 4$-$5  &   220.6 $\pm$     6.9 $\pm$     3.1 $\pm$    15.7  &   199.1 $\pm$     4.7 $\pm$     1.2 $\pm$    14.2  &   185.0 $\pm$     4.6 $\pm$     0.8 $\pm$    13.2 \\
 5$-$6  &   181.2 $\pm$     6.1 $\pm$     4.2 $\pm$    12.9  &   177.4 $\pm$     4.4 $\pm$     1.8 $\pm$    12.6  &   161.4 $\pm$     4.3 $\pm$     0.8 $\pm$    11.5 \\
 6$-$7  &   140.1 $\pm$     5.3 $\pm$     4.5 $\pm$    10.0  &   128.2 $\pm$     3.7 $\pm$     0.9 $\pm$     9.1  &   116.0 $\pm$     3.7 $\pm$     0.6 $\pm$     8.3 \\
 7$-$8  &   106.2 $\pm$     4.6 $\pm$     1.6 $\pm$     7.6  &   106.3 $\pm$     3.4 $\pm$     1.5 $\pm$     7.6  &    93.9 $\pm$     3.3 $\pm$     1.2 $\pm$     6.7 \\
 8$-$9  &    75.3 $\pm$     3.8 $\pm$     2.3 $\pm$     5.4  &    74.0 $\pm$     2.8 $\pm$     0.7 $\pm$     5.3  &    62.3 $\pm$     2.6 $\pm$     0.5 $\pm$     4.4 \\
 9$-$10  &    60.9 $\pm$     3.3 $\pm$     1.5 $\pm$     4.3  &    56.7 $\pm$     2.4 $\pm$     0.7 $\pm$     4.0  &    51.1 $\pm$     2.4 $\pm$     0.4 $\pm$     3.6 \\
 10$-$11  &    51.4 $\pm$     3.0 $\pm$     1.7 $\pm$     3.7  &    44.4 $\pm$     2.1 $\pm$     0.8 $\pm$     3.2  &    32.9 $\pm$     1.9 $\pm$     0.3 $\pm$     2.3 \\
 11$-$12  &    34.7 $\pm$     2.5 $\pm$     1.4 $\pm$     2.5  &    31.1 $\pm$     1.7 $\pm$     0.6 $\pm$     2.2  &    25.6 $\pm$     1.6 $\pm$     0.3 $\pm$     1.8 \\
 12$-$13  &    29.7 $\pm$     2.2 $\pm$     0.3 $\pm$     2.1  &    21.5 $\pm$     1.4 $\pm$     0.9 $\pm$     1.5  &    17.0 $\pm$     1.3 $\pm$     0.2 $\pm$     1.2 \\
 13$-$14  &    22.3 $\pm$     1.9 $\pm$     0.5 $\pm$     1.6  &    16.8 $\pm$     1.3 $\pm$     0.1 $\pm$     1.2  &    12.7 $\pm$     1.1 $\pm$     0.2 $\pm$     0.9 \\
 14$-$15  &    15.6 $\pm$     1.6 $\pm$     0.8 $\pm$     1.1  &    13.8 $\pm$     1.1 $\pm$     0.5 $\pm$     1.0  &    10.4 $\pm$     1.0 $\pm$     0.2 $\pm$     0.7 \\
\hline 
& $3.5<y<4.0$ & $4.0<y<4.5$\\ \hline 
 0$-$1  &    52.0 $\pm$     3.2 $\pm$     0.8 $\pm$     3.7  &    39.7 $\pm$     5.8 $\pm$     0.4 $\pm$     2.8 \\
 1$-$2  &   128.0 $\pm$     4.9 $\pm$     0.9 $\pm$     9.1  &    82.6 $\pm$     8.0 $\pm$     1.9 $\pm$     5.9 \\
 2$-$3  &   167.1 $\pm$     5.5 $\pm$     2.2 $\pm$    11.9  &   125.4 $\pm$     9.6 $\pm$     1.3 $\pm$     8.9 \\
 3$-$4  &   166.4 $\pm$     5.5 $\pm$     1.2 $\pm$    11.8  &   130.2 $\pm$     9.3 $\pm$     1.2 $\pm$     9.3 \\
 4$-$5  &   148.9 $\pm$     5.2 $\pm$     1.3 $\pm$    10.6  &    95.6 $\pm$     7.6 $\pm$     1.1 $\pm$     6.8 \\
 5$-$6  &   101.2 $\pm$     4.2 $\pm$     0.5 $\pm$     7.2  &    68.4 $\pm$     6.2 $\pm$     0.6 $\pm$     4.9 \\
 6$-$7  &    79.1 $\pm$     3.7 $\pm$     0.5 $\pm$     5.6  &    58.5 $\pm$     5.6 $\pm$     0.5 $\pm$     4.2 \\
 7$-$8  &    63.9 $\pm$     3.3 $\pm$     0.6 $\pm$     4.5  &    41.2 $\pm$     4.7 $\pm$     0.5 $\pm$     2.9 \\
 8$-$9  &    47.0 $\pm$     2.8 $\pm$     0.3 $\pm$     3.3  &    30.5 $\pm$     3.9 $\pm$     0.6 $\pm$     2.2 \\
 9$-$10  &    34.8 $\pm$     2.4 $\pm$     0.3 $\pm$     2.5  &    16.5 $\pm$     2.8 $\pm$     0.2 $\pm$     1.2 \\
 10$-$11  &    23.5 $\pm$     2.0 $\pm$     0.2 $\pm$     1.7  &    13.4 $\pm$     2.4 $\pm$     0.2 $\pm$     1.0 \\
 11$-$12  &    17.8 $\pm$     1.7 $\pm$     0.2 $\pm$     1.3  &    12.1 $\pm$     2.3 $\pm$     0.2 $\pm$     0.9 \\
 12$-$13  &    13.4 $\pm$     1.4 $\pm$     0.2 $\pm$     1.0  &     6.6 $\pm$     1.7 $\pm$     0.1 $\pm$     0.5 \\
 13$-$14  &    10.0 $\pm$     1.2 $\pm$     0.2 $\pm$     0.7  &     3.4 $\pm$     1.1 $\pm$     0.1 $\pm$     0.2 \\
 14$-$15  &     7.2 $\pm$     1.0 $\pm$     0.1 $\pm$     0.5  &     0.7 $\pm$     0.5 $\pm$     0.0 $\pm$     0.1 \\

\bottomrule
\end{tabular}
}
\end{center}
\end{table}

\renewcommand{\arraystretch}{1.35}
\begin{table}[!ht]
\begin{center}
\caption{\small \label{tab::xs_twos} Double-differential production cross-sections $\frac{{\rm d}^2\sigma}{{\rm d}p_{\rm T}{\rm d}y}\times \mathcal{B}^{2S}$ in pb/(\gevc) for the \twos\ meson in 
bins of  transverse momentum and rapidity, assuming no polarisation. 
The first error is statistical, the second is the component of the systematic uncertainty that is uncorrelated between bins and the third is the correlated component.
Regions where the number of events was not large enough to perform a measurement are indicated with a dash.
}
\scalebox{0.8}{
\begin{tabular}{cccc}
\\
\midrule
$\pt\,(\gevc)$ 
& $2.0<y<2.5$ 
& $2.5<y<3.0$ 
& $3.0<y<3.5$ \\
\midrule
 0$-$1  &    14.5 $\pm$     1.8 $\pm$     0.3 $\pm$     1.0  &    15.2 $\pm$     1.3 $\pm$     0.7 $\pm$     1.1  &    11.7 $\pm$     1.1 $\pm$     0.1 $\pm$     0.8 \\
 1$-$2  &    33.2 $\pm$     2.7 $\pm$     1.3 $\pm$     2.4  &    42.6 $\pm$     2.1 $\pm$     0.3 $\pm$     3.1  &    39.0 $\pm$     2.1 $\pm$     0.2 $\pm$     2.8 \\
 2$-$3  &    49.3 $\pm$     3.3 $\pm$     1.0 $\pm$     3.5  &    54.3 $\pm$     2.4 $\pm$     1.0 $\pm$     3.9  &    45.0 $\pm$     2.2 $\pm$     0.2 $\pm$     3.2 \\
 3$-$4  &    56.8 $\pm$     3.5 $\pm$     0.6 $\pm$     4.1  &    61.0 $\pm$     2.6 $\pm$     0.4 $\pm$     4.4  &    48.2 $\pm$     2.3 $\pm$     0.2 $\pm$     3.5 \\
 4$-$5  &    46.4 $\pm$     3.1 $\pm$     0.5 $\pm$     3.3  &    45.1 $\pm$     2.2 $\pm$     0.5 $\pm$     3.2  &    39.5 $\pm$     2.1 $\pm$     0.3 $\pm$     2.8 \\
 5$-$6  &    51.1 $\pm$     3.2 $\pm$     0.8 $\pm$     3.7  &    48.9 $\pm$     2.3 $\pm$     0.3 $\pm$     3.5  &    38.1 $\pm$     2.1 $\pm$     0.6 $\pm$     2.7 \\
 6$-$7  &    43.4 $\pm$     2.9 $\pm$     0.9 $\pm$     3.1  &    34.9 $\pm$     1.9 $\pm$     0.6 $\pm$     2.5  &    29.3 $\pm$     1.8 $\pm$     0.2 $\pm$     2.1 \\
 7$-$8  &    28.6 $\pm$     2.4 $\pm$     0.6 $\pm$     2.1  &    29.3 $\pm$     1.8 $\pm$     0.7 $\pm$     2.1  &    25.5 $\pm$     1.7 $\pm$     0.3 $\pm$     1.8 \\
 8$-$9  &    21.2 $\pm$     2.0 $\pm$     0.1 $\pm$     1.5  &    23.7 $\pm$     1.6 $\pm$     0.2 $\pm$     1.7  &    17.3 $\pm$     1.4 $\pm$     0.1 $\pm$     1.2 \\
 9$-$10  &    20.9 $\pm$     2.0 $\pm$     0.2 $\pm$     1.5  &    16.1 $\pm$     1.3 $\pm$     0.2 $\pm$     1.2  &    13.2 $\pm$     1.2 $\pm$     0.1 $\pm$     0.9 \\
 10$-$11  &    16.0 $\pm$     1.7 $\pm$     0.2 $\pm$     1.1  &    13.4 $\pm$     1.2 $\pm$     0.2 $\pm$     1.0  &    13.4 $\pm$     1.2 $\pm$     0.2 $\pm$     1.0 \\
 11$-$12  &    14.9 $\pm$     1.6 $\pm$     0.2 $\pm$     1.1  &    10.2 $\pm$     1.0 $\pm$     0.2 $\pm$     0.7  &    10.1 $\pm$     1.0 $\pm$     0.1 $\pm$     0.7 \\
 12$-$13  &     7.8 $\pm$     1.2 $\pm$     0.2 $\pm$     0.6  &    10.4 $\pm$     1.0 $\pm$     0.2 $\pm$     0.7  &     8.4 $\pm$     1.0 $\pm$     0.2 $\pm$     0.6 \\
 13$-$14  &     5.6 $\pm$     1.0 $\pm$     0.2 $\pm$     0.4  &     7.0 $\pm$     0.8 $\pm$     0.1 $\pm$     0.5  &     5.4 $\pm$     0.8 $\pm$     0.1 $\pm$     0.4 \\
 14$-$15  &     4.0 $\pm$     0.8 $\pm$     0.3 $\pm$     0.3  &     5.8 $\pm$     0.8 $\pm$     0.1 $\pm$     0.4  &     3.5 $\pm$     0.6 $\pm$     0.1 $\pm$     0.3 \\
\hline 
& $3.5<y<4.0$ & $4.0<y<4.5$\\ \hline 
 0$-$1  &    11.2 $\pm$     1.5 $\pm$     0.1 $\pm$     0.8  & -- \\
 1$-$2  &    26.3 $\pm$     2.2 $\pm$     0.2 $\pm$     1.9  & -- \\
 2$-$3  &    35.0 $\pm$     2.5 $\pm$     0.3 $\pm$     2.5  & -- \\
 3$-$4  &    44.7 $\pm$     2.9 $\pm$     0.3 $\pm$     3.2  & -- \\
 4$-$5  &    40.6 $\pm$     2.7 $\pm$     0.2 $\pm$     2.9  &    14.5 $\pm$     2.9 $\pm$     0.1 $\pm$     1.0 \\
 5$-$6  &    28.4 $\pm$     2.3 $\pm$     0.1 $\pm$     2.0  &    12.8 $\pm$     2.8 $\pm$     0.1 $\pm$     0.9 \\
 6$-$7  &    19.8 $\pm$     1.8 $\pm$     0.1 $\pm$     1.4  &    19.5 $\pm$     3.2 $\pm$     0.3 $\pm$     1.4 \\
 7$-$8  &    20.1 $\pm$     1.9 $\pm$     0.1 $\pm$     1.4  &    17.0 $\pm$     3.0 $\pm$     0.2 $\pm$     1.2 \\
 8$-$9  &    14.8 $\pm$     1.6 $\pm$     0.3 $\pm$     1.1  &     8.7 $\pm$     2.1 $\pm$     0.1 $\pm$     0.6 \\
 9$-$10  &     8.7 $\pm$     1.2 $\pm$     0.1 $\pm$     0.6  & -- \\
 10$-$11  &     6.5 $\pm$     1.1 $\pm$     0.2 $\pm$     0.5  & -- \\
 11$-$12  &     7.0 $\pm$     1.1 $\pm$     0.1 $\pm$     0.5  & -- \\
 12$-$13  &     2.7 $\pm$     0.7 $\pm$     0.1 $\pm$     0.2  & -- \\
 13$-$14  &     4.3 $\pm$     0.8 $\pm$     0.1 $\pm$     0.3  & -- \\
 14$-$15  & --  & -- \\

\bottomrule
\end{tabular}
}
\end{center}
\end{table}

\renewcommand{\arraystretch}{1.35}
\begin{table}[!ht]
\begin{center}
\caption{\small \label{tab::xs_threes} Double-differential production cross-sections $\frac{{\rm d}^2\sigma}{{\rm d}p_{\rm T}{\rm d}y}\times \mathcal{B}^{3S}$ in pb/(\gevc) for the \threes\ meson in 
bins of transverse momentum and rapidity, assuming no polarisation. 
The first error is statistical, the second is the component of the systematic uncertainty that is uncorrelated between bins and the third is the correlated component.
Regions where the number of events was not large enough to perform a measurement are indicated with a dash.
}
\scalebox{0.8}{
\begin{tabular}{cccc}
\\
\midrule
$\pt\,(\gevc)$ 
& $2.0<y<2.5$ 
& $2.5<y<3.0$ 
& $3.0<y<3.5$ \\
\midrule
 0$-$1  &     4.8 $\pm$     1.0 $\pm$     0.1 $\pm$     0.4  &     5.5 $\pm$     0.8 $\pm$     0.1 $\pm$     0.4  &     4.7 $\pm$     0.7 $\pm$     0.1 $\pm$     0.3 \\
 1$-$2  &    17.9 $\pm$     2.0 $\pm$     0.3 $\pm$     1.3  &    18.2 $\pm$     1.4 $\pm$     0.2 $\pm$     1.3  &    17.3 $\pm$     1.4 $\pm$     0.1 $\pm$     1.3 \\
 2$-$3  &    25.5 $\pm$     2.3 $\pm$     0.4 $\pm$     1.9  &    17.3 $\pm$     1.3 $\pm$     0.1 $\pm$     1.3  &    21.3 $\pm$     1.5 $\pm$     0.1 $\pm$     1.6 \\
 3$-$4  &    22.9 $\pm$     2.2 $\pm$     0.3 $\pm$     1.7  &    26.2 $\pm$     1.7 $\pm$     0.2 $\pm$     1.9  &    22.8 $\pm$     1.6 $\pm$     0.1 $\pm$     1.7 \\
 4$-$5  &    23.6 $\pm$     2.2 $\pm$     0.3 $\pm$     1.7  &    18.0 $\pm$     1.4 $\pm$     0.1 $\pm$     1.3  &    19.9 $\pm$     1.5 $\pm$     0.1 $\pm$     1.5 \\
 5$-$6  &    19.6 $\pm$     2.0 $\pm$     0.3 $\pm$     1.4  &    21.8 $\pm$     1.5 $\pm$     0.2 $\pm$     1.6  &    18.4 $\pm$     1.4 $\pm$     0.1 $\pm$     1.4 \\
 6$-$7  &    21.4 $\pm$     2.1 $\pm$     0.4 $\pm$     1.6  &    18.9 $\pm$     1.4 $\pm$     0.2 $\pm$     1.4  &    19.8 $\pm$     1.5 $\pm$     0.2 $\pm$     1.5 \\
 7$-$8  &    13.7 $\pm$     1.6 $\pm$     0.3 $\pm$     1.0  &    16.1 $\pm$     1.3 $\pm$     0.2 $\pm$     1.2  &    12.0 $\pm$     1.2 $\pm$     0.2 $\pm$     0.9 \\
 8$-$9  &    14.2 $\pm$     1.6 $\pm$     0.6 $\pm$     1.1  &    12.5 $\pm$     1.1 $\pm$     0.2 $\pm$     0.9  &     9.3 $\pm$     1.0 $\pm$     0.1 $\pm$     0.7 \\
 9$-$10  &     8.7 $\pm$     1.3 $\pm$     0.3 $\pm$     0.6  &    10.4 $\pm$     1.0 $\pm$     0.2 $\pm$     0.8  &    10.8 $\pm$     1.1 $\pm$     0.1 $\pm$     0.8 \\
 10$-$11  &     9.6 $\pm$     1.3 $\pm$     0.5 $\pm$     0.7  &     7.4 $\pm$     0.9 $\pm$     0.3 $\pm$     0.5  &     6.8 $\pm$     0.8 $\pm$     0.1 $\pm$     0.5 \\
 11$-$12  &     8.1 $\pm$     1.2 $\pm$     0.1 $\pm$     0.6  &     5.9 $\pm$     0.8 $\pm$     0.2 $\pm$     0.4  &     4.0 $\pm$     0.7 $\pm$     0.1 $\pm$     0.3 \\
 12$-$13  &     6.3 $\pm$     1.0 $\pm$     0.1 $\pm$     0.5  &     5.3 $\pm$     0.7 $\pm$     0.1 $\pm$     0.4  &     3.0 $\pm$     0.6 $\pm$     0.0 $\pm$     0.2 \\
 13$-$14  &    10.0 $\pm$     1.3 $\pm$     0.2 $\pm$     0.7  &     4.6 $\pm$     0.7 $\pm$     0.2 $\pm$     0.3  &     3.7 $\pm$     0.6 $\pm$     0.1 $\pm$     0.3 \\
 14$-$15  &     4.0 $\pm$     0.8 $\pm$     0.1 $\pm$     0.3  &     2.2 $\pm$     0.5 $\pm$     0.1 $\pm$     0.2  &     3.5 $\pm$     0.6 $\pm$     0.1 $\pm$     0.3 \\
\hline 
& $3.5<y<4.0$ & $4.0<y<4.5$\\ \hline 
 0$-$1  &     3.3 $\pm$     0.8 $\pm$     0.0 $\pm$     0.2  & -- \\
 1$-$2  &    19.1 $\pm$     1.9 $\pm$     0.2 $\pm$     1.4  & -- \\
 2$-$3  &    20.0 $\pm$     1.9 $\pm$     0.2 $\pm$     1.5  & -- \\
 3$-$4  &    18.6 $\pm$     1.8 $\pm$     0.1 $\pm$     1.4  & -- \\
 4$-$5  &    18.5 $\pm$     1.8 $\pm$     0.1 $\pm$     1.4  &     6.3 $\pm$     2.0 $\pm$     0.1 $\pm$     0.5 \\
 5$-$6  &    11.0 $\pm$     1.4 $\pm$     0.1 $\pm$     0.8  &     5.3 $\pm$     1.7 $\pm$     0.1 $\pm$     0.4 \\
 6$-$7  &     9.3 $\pm$     1.3 $\pm$     0.1 $\pm$     0.7  &     1.0 $\pm$     0.7 $\pm$     0.0 $\pm$     0.1 \\
 7$-$8  &     8.1 $\pm$     1.2 $\pm$     0.1 $\pm$     0.6  &     7.3 $\pm$     1.9 $\pm$     0.1 $\pm$     0.5 \\
 8$-$9  &     8.0 $\pm$     1.2 $\pm$     0.1 $\pm$     0.6  &     3.0 $\pm$     1.2 $\pm$     0.0 $\pm$     0.2 \\
 9$-$10  &     7.9 $\pm$     1.1 $\pm$     0.1 $\pm$     0.6  & -- \\
 10$-$11  &     3.6 $\pm$     0.8 $\pm$     0.0 $\pm$     0.3  & -- \\
 11$-$12  &     5.1 $\pm$     0.9 $\pm$     0.1 $\pm$     0.4  & -- \\
 12$-$13  &     4.5 $\pm$     0.9 $\pm$     0.1 $\pm$     0.3  & -- \\
 13$-$14  &     0.7 $\pm$     0.3 $\pm$     0.0 $\pm$     0.1  & -- \\
 14$-$15  & --  & -- \\

\bottomrule
\end{tabular}
}
\end{center}
\end{table}

\renewcommand{\arraystretch}{1.4}
\begin{table}[!ht] 
\begin{center} 
\caption{\small{
   Ratios of cross-sections
  $\twosmm$ and $\threesmm$  with respect to \nobreak{$\onesmm$}
 as a function of $\pt$ in the
range $2.0<y<4.0$, assuming no polarisation. 
The first error is statistical, the second  
is the component of the systematic uncertainty that is uncorrelated between bins and the third is the correlated 
component.
}}\label{tab::ratiospt} 
\resizebox{0.85\textwidth}{!}{ 
\begin{tabular}{ccc} 
\midrule
 $\pt$ (\gevc) & $R^{2S/1S}$ & $R^{3S/1S}$ \\ 
\midrule
   0--1  &   0.210 $\pm$   0.013 $\pm$   0.004 $\pm$   0.006 &   0.075 $\pm$   0.007 $\pm$   0.001 $\pm$   0.003\\ 
   1--2  &   0.219 $\pm$   0.008 $\pm$   0.002 $\pm$   0.006 &   0.109 $\pm$   0.005 $\pm$   0.001 $\pm$   0.004\\ 
   2--3  &   0.219 $\pm$   0.007 $\pm$   0.002 $\pm$   0.006 &   0.096 $\pm$   0.004 $\pm$   0.001 $\pm$   0.003\\ 
   3--4  &   0.244 $\pm$   0.008 $\pm$   0.002 $\pm$   0.007 &   0.107 $\pm$   0.004 $\pm$   0.001 $\pm$   0.004\\ 
   4--5  &   0.226 $\pm$   0.008 $\pm$   0.001 $\pm$   0.006 &   0.104 $\pm$   0.005 $\pm$   0.001 $\pm$   0.004\\ 
   5--6  &   0.267 $\pm$   0.009 $\pm$   0.003 $\pm$   0.007 &   0.116 $\pm$   0.005 $\pm$   0.001 $\pm$   0.004\\ 
   6--7  &   0.270 $\pm$   0.011 $\pm$   0.004 $\pm$   0.008 &   0.151 $\pm$   0.007 $\pm$   0.001 $\pm$   0.005\\ 
   7--8  &   0.280 $\pm$   0.012 $\pm$   0.003 $\pm$   0.008 &   0.137 $\pm$   0.007 $\pm$   0.001 $\pm$   0.005\\ 
   8--9  &   0.298 $\pm$   0.015 $\pm$   0.003 $\pm$   0.008 &   0.170 $\pm$   0.010 $\pm$   0.002 $\pm$   0.006\\ 
   9--10  &   0.280 $\pm$   0.016 $\pm$   0.003 $\pm$   0.008 &   0.190 $\pm$   0.012 $\pm$   0.002 $\pm$   0.007\\ 
  10--11  &   0.330 $\pm$   0.021 $\pm$   0.005 $\pm$   0.009 &   0.181 $\pm$   0.013 $\pm$   0.004 $\pm$   0.006\\ 
  11--12  &   0.379 $\pm$   0.027 $\pm$   0.006 $\pm$   0.011 &   0.203 $\pm$   0.017 $\pm$   0.003 $\pm$   0.007\\ 
  12--13  &   0.401 $\pm$   0.032 $\pm$   0.007 $\pm$   0.011 &   0.235 $\pm$   0.021 $\pm$   0.003 $\pm$   0.008\\ 
  13--14  &   0.388 $\pm$   0.036 $\pm$   0.006 $\pm$   0.011 &   0.292 $\pm$   0.027 $\pm$   0.004 $\pm$   0.010\\ 
\hline 
\end{tabular}
}
\end{center} 
\end{table}

\begin{table}[!ht] 
\begin{center} 
\caption{\small{
   Ratios of cross-sections
  $\twosmm$ and $\threesmm$  with respect to \nobreak{$\onesmm$}
 as a function of $y$ in the
range $2.0<y<4.0$, assuming no polarisation. 
The first error is statistical, the second  
is the component of the systematic uncertainty that is uncorrelated between bins and the third is the correlated 
component.
}}\label{tab::ratiosy} 
\resizebox{0.9\textwidth}{!}{ 
\begin{tabular}{ccc} 
\midrule
 $y$ & $R^{2S/1S}$ & $R^{3S/1S}$ \\ 
\hline 
  2.0--2.5  &   0.2545 $\pm$   0.0068 $\pm$   0.0021 $\pm$   0.0071 &   0.1260 $\pm$   0.0041 $\pm$   0.0008 $\pm$   0.0044\\ 
  2.5--3.0  &   0.2653 $\pm$   0.0050 $\pm$   0.0014 $\pm$   0.0075 &   0.1210 $\pm$   0.0029 $\pm$   0.0005 $\pm$   0.0043\\ 
  3.0--3.5  &   0.2476 $\pm$   0.0051 $\pm$   0.0009 $\pm$   0.0070 &   0.1251 $\pm$   0.0032 $\pm$   0.0004 $\pm$   0.0044\\ 
  3.5--4.0  &   0.2558 $\pm$   0.0075 $\pm$   0.0011 $\pm$   0.0072 &   0.1315 $\pm$   0.0048 $\pm$   0.0004 $\pm$   0.0046\\ 
\hline 
\end{tabular}
}
\end{center} 
\end{table}

\addcontentsline{toc}{section}{References}
\clearpage
\bibliographystyle{LHCb}
\bibliography{ref,quarkonia,main,LHCb-PAPER}
\clearpage
\end{document}